\documentclass[12pt]{article}
\usepackage{geometry}
\usepackage{a4}
\usepackage{graphicx}
\usepackage{epsf}
\usepackage{amsmath}
\usepackage{amssymb}
\usepackage[T1]{fontenc}
\usepackage[utf8]{inputenc}
\usepackage{tabularx,ragged2e,booktabs,caption}
\usepackage{cite}
\newcommand{\be}{\begin{equation}}
\newcommand{\ee}{\end{equation}}

\newcommand{\Rmnum}[1]{\expandafter\@slowromancap\romannumeral #1@}
\newcommand{\bea}{\begin{eqnarray}}
\newcommand{\eea}{\end{eqnarray}}

\begin{document}
%%%%%%%%%%%%%%%%%%%%%%%%%%%%%%%
\def\A{{\mathbb{A}}}
\def\B{{\mathbb{B}}}
\def\C{{\mathbb{C}}}
\def\R{{\mathbb{R}}}
\def\s{{\mathbb{S}}}
\def\T{{\mathbb{T}}}
\def\Z{{\mathbb{Z}}}
\def\W{{\mathbb{W}}}
%%%%%%%%%%%%%%%%%%%%%%%%%%%%
\begin{titlepage}
\title{Very General Holographic Superconductors and Entanglement Thermodynamics}
\author{}
\date{% authors are dated
Anshuman Dey, Subhash Mahapatra, Tapobrata Sarkar
\thanks{\noindent E-mail:~ deyanshu, subhmaha, tapo @iitk.ac.in}
\vskip0.4cm
{\sl Department of Physics, \\
Indian Institute of Technology,\\
Kanpur 208016, \\
India}}
\maketitle
\abstract{We construct and analyze holographic superconductors with generalized higher derivative couplings, in single R-charged black hole backgrounds in four and five dimensions. 
These systems, which we call very general holographic superconductors, have multiple tuning parameters and are shown to exhibit a rich phase structure. We establish the phase diagram
numerically as well as by computing the free energy, and then validated the results by calculating the entanglement entropy for these systems. The entanglement entropy is shown to be a 
perfect indicator of the phase diagram. The differences in the nature of the entanglement entropy in R-charged backgrounds compared to the AdS-Schwarzschild cases are pointed out. 
We also compute the analogue of the entangling temperature for a subclass of these systems and compare the results with non-hairy backgrounds.}
\end{titlepage}

\section{Introduction}

It is by now well recognized that holographic AdS/CFT duality \cite{Maldacena} can provide valuable insights into the physics of strongly coupled systems which may not
be amenable to a perturbative analysis. This duality, which relates a classical theory of AdS gravity to a conformal field theory in
one lower dimension (living on the boundary of the AdS space) has in particular found important applications in the study of strongly coupled condensed matter systems.
Although it is fair to say that connections to realistic condensed matter physics via the holographic correspondence has so far remained elusive, it is important to explore this
line of research and further our understanding towards the ultimate goal of connecting to experimental results.

Two of the most important aspects that have received wide attention in the context of the AdS/CFT correspondence are holographic superconductors, initiated by the works of
\cite{Hartnoll}, \cite{Johnson1}, \cite{Hartnoll_1} and holographic entanglement entropy (HEE), introduced in \cite{Ryu}.
While we expect that the former might capture important physical effects in realistic superconducting systems, the latter should be of importance in, for example, areas related to information theory.
Several authors have, in the recent past, studied various aspects of holographic superconductors and in particular, HEE in that setting \cite{Johnson}.

In a previous work \cite{Dey}, we had built a model of a generalized holographic superconductor, with a generalized form of higher derivative couplings, following the work
of \cite{Garcia}, \cite{Kuang}. To our knowledge this is the most general phenomenological model of a holographic superconductor constructed till date, and shows rich phase structure
compared to other models considered in the literature. We call such a model (to be elaborated upon in sequel) a very general holographic superconductor (VGHS).
The work of \cite{Dey} dealt with such models in the background of an AdS-Schwarzschild black hole. One of the main purposes of the present paper is to construct such models
of holographic superconductors in the background of planar single R-charged black hole solutions including backreaction, and to study features of their HEE.
In this introductory section, we will provide a brief overview of the topics to be covered in the rest of the paper and then proceed to summarize
our main results.

In the simplest realization of a holographic superconductor, it was shown in \cite{Gubser} that AdS black holes with Abelian Higgs matter become unstable to forming scalar hair near the
horizon, below a certain critical temperature $T_c$. The main reason for this instability is the presence of a minimal coupling between the scalar and the gauge field, which can make the effective mass term
of the scalar field sufficiently negative near the horizon. In the dual boundary field theory, this complex scalar field instability corresponds to a non zero vacuum expectation value (VEV) of the charged scalar
operator which is dual to the scalar field in the bulk \cite{Hartnoll}. The non zero VEV of the scalar operator corresponds to a spontaneous breaking of the $U(1)$ gauge symmetry and therefore indicates
a phase transition from a normal to a superconducting phase, with the scalar operator playing the role of an order parameter.
Strictly speaking, at the boundary, it is a global $U(1)$ symmetry which is broken spontaneously
and therefore these models more properly describe a holographic charged superfluid. However, one can weakly gauge this symmetry and can still describe superconductivity \cite{Hartnoll_1}.
Indeed, it was shown explicitly in \cite{Hartnoll} that the DC conductivity is infinite in these models, which is one of the main characteristic properties of superconductors. 
Other important features of superconductivity, such as the the existence of an energy gap can also be shown in the context of holographic 
superconductors \cite{Hartnoll_1}. Specifically, in \cite{Roberts}, a universal
ratio of $\omega_g/T_c\sim 8$, where $\omega_g$ is the gap in the frequency dependent conductivity and $T_c$ the critical temperature, was found. 
A gap in the optical conductivity implies an energy gap in the charge spectrum,
which is, as mentioned before, an essential feature of superconductivity. In the weakly coupled BCS theory, $\omega_g$ can also be thought as the energy required to break a Cooper pair into its
constitutive electrons. Prediction of this ratio from holography, which is twice compared to the BCS theory, indicates the strongly interacting nature of the boundary theory, although
by now a large number of exceptions to this result are also known \cite{Soda}\cite{Pan}. Also, Meissner type effects 
can be shown to exist in holographic superconductors \cite{Johnson1}, \cite{Hartnoll_1}.

The original model of holographic superconductors was subsequently generalised in \cite{Hartnoll_1} to include the effects of backreaction of the Abelian Higgs matter fields on gravity.
Here, it was argued that effects of backreaction do not change the physics too much, and that essentially all the main results are captured by the probe limit.
However, there are a few differences and in particular, it was found that the effects of backreaction generally make the condensation harder to form.

An important generalization of the original model of holographic superconductors was considered in \cite{Garcia}, where the $U(1)$ symmetry in the boundary is broken by a St\"{u}ckelberg
mechanism. These models have subsequently been called generalized holographic superconductors in the literature.
The essential idea here is to consider a non-minimal coupling between the scalar and the gauge field in a gauge invariant way. The importance of this models lies in the fact that one can tune
the order of the phase transition by introducing additional parameters in the theory, which might be important in realistic systems. With one such parameter, interestingly, several authors
found the existence of a first order phase transition from the normal to the superconducting phase, and a metastable region in the superconducting phase\cite{Cai}\cite{Subhash}.
 \footnote{See \cite{Yan} for a treatment of generalized superconductor with backreaction effects.}
This is phenomenologically important, since there are a number of superconductors which show first order phase transitions \cite{Bianchi}\cite{Yonezawa}. We should of course emphasize that
the first order phase transitions in holographic superconductors are typically studied in the absence of external magnetic fields, unlike real superconductors. However, the issue of
of phase transitions in superconductors continue to be an important topic for research in condensed matter physics, and first order transitions inside the superconducting phase
at zero magnetic field have seen some interesting development of late \cite{Tokiwa}. As of now, it is possibly fair to say that predictions from holographic
superconductors via AdS/CFT are still far from being tested in the laboratory.

The models mentioned in the previous paragraph are phenomenological in nature, in which the fields and the interactions between them are put in by hand, without actually 
deriving them from consistent truncations of a string theory, i.e this is a bottom-up approach. In such an approach, the full microscopic discerption of holographic superconductors 
(like we have for BCS superconductors) are not known.
To have such a microscopic description, one has to embed the theory in a string theory, i.e follow a top-down approach (see e.g \cite{Bobev}), which might be substantially
more complicated than a bottom-up one, which is the viewpoint we take in this paper. In this bottom-up description, a model with higher derivative interactions of the the scalar and the gauge field via a
coupling constant $\eta$ was proposed in \cite{Kuang}. These authors analysed the formation of droplets in an external magnetic field in the probe limit and subsequently also found some signature of
the ``proximity effect''  \cite{Kuang_1}.  A non-trivial generalization of the model of \cite{Kuang} was considered in \cite{Dey}, by introducing two analytical functions of the scalar field in a gauge invariant way.
The usefulness of this latter model, which we have called a very general holographic superconductor, lies in the fact that one has multiple tunable parameters in the theory,
which provides a far richer phase structure compared to minimally coupled holographic superconductors. For two such parameters, an exotic ``window'' of first order phase transitions from
the normal to the superconducting phase was found in \cite{Dey}. It is certainly not clear how this might be related to current experimental observations, but
if in future, evidence for existence of such systems are found, the VGHS might provide a strong coupling realization of the same.

Now we turn to the concept of entanglement entropy, which has also received a lot of attention of late, and is considered in the later part of this paper. Qualitatively speaking,
if a quantum system is divided into two subsystems $\mathcal{A}$ and $\mathcal{B}$, measurements on $\mathcal{A}$ will affect those on $\mathcal{B}$, if the two subsystems are entangled.
Entanglement entropy (EE) is a quantitative measure which tells us how strongly these two subsystems are entangled or correlated.
Since EE is related to the degrees of freedom of the system, in condensed matter physics it is an important tool to quantify the appearance of a phase transition, as well as its order.
However, it is difficult to calculate the EE of a quantum field theory beyond $1+1$ dimensions. This problem was bypassed by Ryu and Takayanagi \cite{Ryu}, who proposed a simple formula to calculate
the EE in the holographic scenario, which is now referred to as the holographic entanglement entropy. Several computations of the HEE have been done using the Ryu-Takayanagi prescription
in different contexts, and the results are in good agreement with the standard CFT results.

Recently, \cite{Johnson} used this prescription to study HEE for a ``strip geometry'' (to be elaborated upon later in this paper) in
the context of holographic metal-superconductor phase transitions, and showed that the HEE not only captures the appearance of the phase transition but also its order. 
The results of this paper also show that for a fixed strip geometry, the HEE in the superconducting phase is always less than that in the normal phase, which is in some sense expected, since
below the transition temperature, some of the degrees of freedom get condensed. Subsequent to this, in \cite{Cai}, \cite{Cai2}, analysis of the behavior of the HEE in the context of
holographic insulator-superconductor phase transitions was done. In \cite{Kuang_1}, a holographic superconductor with higher derivative couplings is considered, and 
these authors calculate the entanglement entropy to study the proximity effect in superconductors. Another recent development in the context of HEE is the interesting notion of the
entangling temperature, which first appeared in \cite{Bhattacharya}. In this paper, it was shown that there exists an analogue of the first law of thermodynamics
with the HEE playing the role of the usual entropy. For a small subsystem, the change in HEE is proportional to the change in the energy of the subsystem and
the proportionality constant, which is given by the size of the entangling region, is interpreted as the inverse of the entangling temperature.

%%%%%%%%%%%%%%%%%%%%%%%%%%%%%%%%%%%%%%%%%%%%%%%%%%%%%%%%%%%%%%%%%
\begin{figure}[t!]
\centering
\includegraphics[width=3in,height=2.5in]{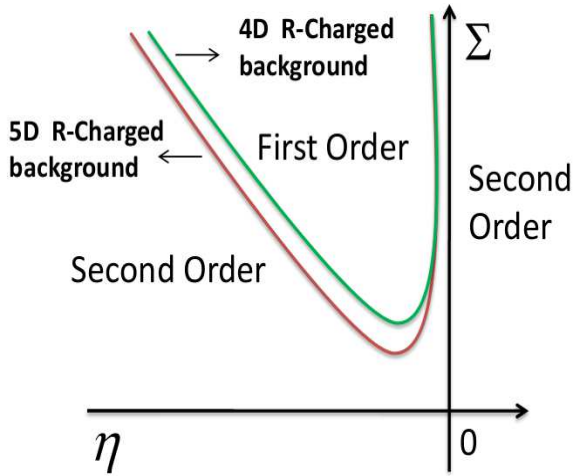}
\caption{Qualitative phase diagram of the VGHS in the parameter space in the probe limit.}
\label{phasediagram}
\end{figure}
%%%%%%%%%%%%%%%%%%%%%%%%%%%%%%%%%%%%%%%%%%%%%%%%%%%%%%%%%%%%%%%%%
Having briefly reviewed known literature on the topic, we now state our intent.
The purpose of the present paper is to extend and complete the study of the VGHS in R-charged backgrounds, in lines with the
discussion above. The organization of this paper and the main results contained herein are summarized below.\\
\noindent
$\bullet$ In section 2, we construct the VGHS in four dimensional planar R-charged black hole backgrounds, including back reaction
effects. We show that in these backgrounds also, there is a window of first order metal - superconductor phase transitions, i.e these first order transitions appear when one appropriately tunes
the parameters of the theory in a certain range. We check this result by establishing the nature of the difference in the free energy between the superconductor and the normal phases.
This is done for R-charged black hole backgrounds in four and five dimensions. A qualitative phase diagram for our VGHS in the probe limit is shown in fig.(\ref{phasediagram}).\\
\noindent
$\bullet$ Section 3 is devoted to the study of the VGHS in five dimensional planar R-charged backgrounds. Since the analysis is qualitatively similar to the one
carried out in section 2, we relegate the details of calculation in this section to Appendix A. Here also we find a window of first order phase transitions within a certain range of parameters of
the theory. Again, this is validated by calculating the free energy. \\
$\bullet$
Next, in section 4, we study holographic entanglement entropy for the R-charged backgrounds studied above. As a warm up exercise, we first calculate the HEE for the VGHS in an
AdS Schwarzschild background, and show that the HEE correctly captures the information about the window of first order phase transitions that we have mentioned.
(The details of the gravity side of this calculation are relegated to Appendix B).
We then perform the analysis for R-charged backgrounds and show that the HEE is again an effective tool to pinpoint the window of first order phase transitions in these cases. However,
unlike other cases studied in the literature, we find that the HEE for four dimensional R-charged background actually seems to increases in the superconducting phases (compared to the normal phase) 
whereas the free energy shows expected behavior. Currently, we do not have a proper physical explanation for this, nevertheless, we will provide some discussions towards the end of this section.
This feature is absent in five dimensional backgrounds. \\
\noindent
$\bullet$
In section 5, we study the entangling temperature for the VGHS, to look for relations similar to the first law of thermodynamics
with HEE. Our method here is numerical, and we fit the metric and the backreaction parameter with appropriate polynomial functions and extract the entangling temperature.
We find some expected variations from the results obtained in \cite{Bhattacharya}. In this section, we confine ourselves to AdS-Schwarzschild backgrounds, and point out
some difficulties of a similar calculation in R-charged examples. Our results broadly indicate the need to understand better aspects of the entangling temperature
for the VGHS in R-charged backgrounds.

Finally, section 6 ends this paper with a summary of our results and possible directions for future research.

\section{4-D R-charged black hole backgrounds}

In this section we will set up a model for the VGHS in four dimensional R-charged backgrounds. This will also serve to illustrate the basic notations and
conventions used in the rest of the paper. We mention in the outset that we will deal with planar R-charged backgrounds with a single charge turned on. With multiple chemical
potentials, the solution seems to be intractable.

Recall that R-charged black holes form the gravity duals to rotating branes in various dimensions. As an example, while the gravity dual to a D3-brane
configuration is ${\rm AdS}_5 \times S^5$, adding spin to the D3-brane configuration in directions orthogonal to its world volume amounts to adding rotations that correspond to a global $SO(6)$ R-symmetry
of the $N=4$ conformal field theory that resides on the brane and is related to the $SO(6)$ symmetry of the $D=5$, $N=8$ gauged supergravity that arises upon a Kaluza Klein reduction
of the spinning brane configuration on $S^5$. The three $U(1)$ gauge charges in the ${\rm AdS}_5$ supergravity are thus related to the spins on the brane world volume, and give rise to three
chemical potentials. In a similar manner, black holes in four dimensional $N=8$ ${\rm AdS}$ supergravity contains four R-charges that correspond to an $SO(8)$ gauge symmetry arising out
of a Kaluza Klein reduction of spinning M2-brane configurations on $S^7$. Holographic superconductors can be built by considering an Abelian Higgs model in these geometries.

For the four dimensional single R-charged black hole, we start with the following action
\begin{eqnarray}
\textit{S} &=& \int \mathrm{d^{4}}x\! \sqrt{-g}\biggl[\frac{1}{2\kappa^{2}}\biggl(R+\frac{3}{L^{2}}\biggl(H^{1/2}+H^{-1/2}\biggr)\biggr)
-\frac{L^2 H^{3/2}}{8}\textit{F}_{\mu\nu}\textit{F}^{\mu\nu}-\frac{3}{8}\frac{(\partial H)^2}{H^2} \nonumber \\ 
&-&\frac{1}{2}|\textit{D}\tilde{\Psi}|^{2} -\frac{1}{2}m^{2}|\tilde{\Psi}|^{2} -\frac{\eta}{2}|\textit{F}_{\mu\nu}\textit{D}^{\nu}\tilde{\Psi}|^{2}\biggl] \,
\label{action}
\end{eqnarray}
Here, $\kappa$ is related to the four dimensional Newton's constant, $L$ is the AdS length scale and  $\tilde{\Psi}$
is a complex scalar field with charge $q$ and mass $m$. Also,
$F=dA$ and $D_{\mu}=\partial_{\mu}-i q A_{\mu}$.
For $\tilde{\Psi}=0$, the above action reduces to that of the single R-charged black hole background (see e.g \cite{natsuume}) with $H(r) = 1 + kr_h/r$, $r_h$ being the horizon radius,
and $k$ a charge parameter.
Also, the last term in Eq.(\ref{action}) describes the higher derivative interaction between the scalar field and the field strength tensor. The form of the interaction can 
be motivated from a Landau-Ginzburg analysis, but we will prefer to study this from a phenomenological point of view. 
Rewriting the charged scalar field $\tilde{\Psi}= \Psi  e^{i \alpha}$, the action can be cast as
\begin{eqnarray}
\textit{S} &=& \int \mathrm{d^{4}}x\! \sqrt{-g}\biggl[\frac{1}{2\kappa^{2}}\biggl(R+\frac{3}{L^{2}}\biggl(H^{1/2}+H^{-1/2}\biggr)\biggr)-
\frac{L^2 H^{3/2}}{8}\textit{F}_{\mu\nu}\textit{F}^{\mu\nu}-\frac{3}{8}\frac{(\partial H)^2}{H^2} \nonumber \\ 
&-&\frac{(\partial_{\mu}\Psi)^2}{2} -\frac{m^2\Psi^2}{2}  -\frac{\eta}{2}\textit{F}_{\mu\nu}\partial^{\nu}\Psi \textit{F}^{\mu\sigma}\partial_{\sigma}\Psi -\frac{\Psi^2(\partial\alpha-q A)^2}{2} \nonumber \\
&-&\frac{\eta}{2}\Psi^2\biggl(\textit{F}^{\mu\nu}(\partial_{\nu}\alpha-q A_{\nu})\biggr)^{2} \biggr] \,
\label{actionpsi2}
\end{eqnarray}
The $U(1)$ symmetry in the above action is now given by $\alpha\rightarrow \alpha+q \lambda$ and $A_{\mu}\rightarrow A_{\mu}+\partial_{\mu}\lambda$.
Following \cite{Garcia},\cite{Dey}, the above action can be generalized in gauge invariant way
\begin{eqnarray}
\textit{S} &=& \int \mathrm{d^{4}}x\! \sqrt{-g}\biggl[\frac{1}{2\kappa^{2}}\biggl(R+\frac{3}{L^{2}}\biggl(H^{1/2}+H^{-1/2}\biggr)\biggr)
-\frac{L^2 H^{3/2}}{8}\textit{F}_{\mu\nu}\textit{F}^{\mu\nu}-\frac{3}{8}\frac{(\partial H)^2}{H^2} \nonumber \\ 
&-&\frac{(\partial_{\mu}\Psi)^2}{2}-\frac{\eta}{2}\textit{F}_{\mu\nu}\partial^{\nu}\Psi \textit{F}^{\mu\sigma}\partial_{\sigma}\Psi  -\frac{m^2\Psi^2}{2} -\frac{|\textrm{G}(\Psi)|(\partial\alpha-q A)^2}{2}\nonumber \\
&-& \frac{\eta}{2}|\textrm{K}(\Psi)|\biggl(\textit{F}^{\mu\nu}(\partial_{\nu}\alpha-q A_{\nu})\biggr)^{2} \biggr]
\label{actionpsi3}
\end{eqnarray}
here $\textrm{G}(\Psi)$ and $\textrm{K}(\Psi)$ are two analytic functions of $\Psi$ whose general form will be specified in subsequent text.
Eq.(\ref{actionpsi3}) defines the VGHS. If $\textrm{K}(\Psi) = \Psi^2$, we will obtain the generalized holographic superconductor of \cite{Garcia} along with a higher derivative
coupling. However, as we show in sequel, more general forms of $\textrm{K}(\Psi)$ leads to a rich phase structure in the theory.  \footnote{In the rest of this paper, we perform the
computations by setting $L=1$ and $q=1$.}

Now for hairy black hole like solutions with backreaction, we consider the following ansatz \footnote{All numerical calculations in this paper are performed using MATHEMATICA routines.
We find that in some situations, a conformally equivalent metric ansatz
$\textit{d}s^{2}=-g(r)e^{-\xi(r)}dt^{2}+r^{2}(dx^{2}+dy^{2})+\frac{dr^{2}}{g(r)}$ reduces the computation time considerably, while giving the same numerical results as when one uses eq.(\ref{metric4D}).}
\begin{eqnarray}
\textit{d}s^{2}=-g(r)H(r)^{-1/2}e^{-\chi(r)}dt^{2}+H(r)^{1/2}r^{2}(dx^{2}+dy^{2})+H(r)^{1/2}\frac{dr^{2}}{g(r)}
\label{metric4D}
\end{eqnarray}
\begin{equation}
\Psi=\Psi(r),~~~A=\Phi(r)dt
\end{equation}
We will henceforth consider a particular gauge where $\alpha =0$. In this gauge, the equation of motion for the scalar field $\Psi$ can be obtained as
\begin{eqnarray}
\Psi '' \left(1-\eta  e^{\chi } \Phi'^2\right)+ \frac{H e^{\chi} \Phi^2}{2g^2}\frac{d\textrm{G}(\Psi)}{d\Psi}  -\frac{\eta H e^{2\chi}\Phi^2 \Phi'^2}{2g^2}
\frac{d\textrm{K}(\Psi)}{d\Psi}+ \Psi '\left( \frac{2}{r}+\frac{g'}{g}-\frac{\chi '}{2}\right) &\nonumber \\  -\frac{m^2 H^{1/2} \Psi}{g} - \eta \Psi' \left(\frac{e^{\chi} g' \Phi '^2}{g}+\frac{
e^{\chi } \Phi'^2 \chi'}{2}+\frac{2 e^{\chi} \Phi'^2}{r}+2 e^{\chi} \Phi' \Phi''\right)=0
\label{psieom4D}
\end{eqnarray}
Similarly, we get the equation of motion for the zeroth component of the gauge field as
\begin{eqnarray}
&& \Phi''\left(1 -\frac{2 \eta e^{\chi} \Phi^2 \textrm{K}(\Psi)}{g H} + \frac{2 \eta g \Psi'^2}{H^{2}} \right) -\Phi
\left(\frac{2\textrm{G}(\Psi)}{g H}+\frac{2 \eta e^{\chi} \Phi'^2 \textrm{K}(\Psi) }{g H}\right) \nonumber \\ &&  +
\Phi' \left(\frac{2\eta g' \Psi'^2}{H^2}+\frac{\eta g \chi' \Psi'^2}{H^2} + \frac{4\eta g \Psi'^2}{r H^2} + \frac{4\eta g \Psi' \Psi''}{H^2} +
\frac{2}{r} + \frac{2 H'}{H}+\frac{\chi '}{2} \right) \nonumber \\ && + 2\eta \textrm{K}(\Psi) \Phi ^2 \Phi' \left(\frac{e^{\chi} g'}{g^2 H}-
\frac{e^{\chi} \textrm{K}(\Psi)'}{g H \textrm{K}(\Psi)}-\frac{3e^{\chi} \chi'}{2g H}-\frac{2e^{\chi} }{r g H} - \frac{e^{\chi} H'}{g H^2}\right) =0
\label{phieom4D}
\end{eqnarray}
Also, the equation of motion for the $H$ field is given by
\begin{equation}
H'' + H' \left(\frac{2}{r}+\frac{g'}{g}-\frac{\chi'}{2}-\frac{H'}{H} \right)+\frac{e^{\chi} H^{3} \Phi'{^2}}{2g}+\frac{2H}{2\kappa^{2}g}(H-1)=0
\label{Heom4D}
\end{equation}
Finally, the $rr$ and the $(tt-rr)$ components of Einstein equation give
\begin{eqnarray}
 g' - g\chi' +\frac{g}{r}-\frac{3r}{2}(H+1)+\frac{r g H'}{4 H}\left(\frac{g'}{g}-\frac{H'}{4 H}-\chi' \right) && \nonumber \\
 +2\kappa^{2} r \biggl(-\frac{H e^{\chi}\Phi^2\textrm{G}(\Psi)}{4 g} + \frac{H^{1/2}m^{2} \Psi^{2}}{4} - \frac{3 g H'^{2}}{16 H^2} + \frac{3\eta g e^{\chi}\Phi'^{2}\Psi'^{2}}{4}
 && \nonumber \\ +\frac{e^{\chi}H^{2}\Phi'^{2}}{8} - \frac{\eta H e^{2\chi} \Phi^{2}\Phi'^{2}\textrm{K}(\Psi)}{4 g} -\frac{g \Psi'^{2}}{4} \biggr)=0
\label{rreinsteineom4D}
\end{eqnarray}
\begin{eqnarray}
2\kappa^{2} r \biggl(\frac{H e^{\chi}\Phi^2\textrm{G}(\Psi)}{2 g^{2}}+\frac{3 H'^{2}}{8H^{2}} -\frac{\eta H e^{2\chi}\Phi^{2}\Phi'^{2}\textrm{K}(\Psi)}{2 g^{2}}+\frac{\Psi'^{2}}{2}-\frac{\eta e^{\chi}\Phi'^{2}\Psi'^{2}}{2} \biggr)  
&& \nonumber \\ \chi'+\frac{H'}{H}-\frac{3rH'{2}}{8H^{2}}+\frac{rH'\chi'}{4H}+\frac{rH''}{2H}=0
\label{ttrreinsteineom4D}
\end{eqnarray}
In the above equations, we have explicitly suppressed the radial dependence of our variables, and the prime denotes a derivative with respect to the radial coordinate $r$.
Let us  record the expression for the Hawking temperature of the black hole with the geometry in equation (\ref{metric4D}), which is given by
\begin{equation}
T_{H}=\frac{g'(r)e^{-\chi(r)/2}}{4\pi \sqrt{H(r)}}|_{r=r_{h}}
\end{equation}
where $r_{h}$, the radius of the event horizon, is given by the solution of $g(r_{h})=0$. Finally therefore, we have five coupled differential equations which need to be
solved with appropriate boundary conditions. We impose the regularity conditions for $\Phi$ and $\Psi$ at the horizon
\begin{equation}
\Phi(r_{h})=0 ,\ \ \Psi'(r_{h})=\frac{m^2 \sqrt{H(r_{h})}\Psi(r_{h})}{g'(r_{h})\left(1-\eta e^{\chi(r_{h})}\Phi'^{2}(r_{h})\right)}.
\label{horizon behavior4D}
\end{equation}
Near the boundary these fields asymptote to the following expressions
\begin{eqnarray}
\Phi=\mu-\frac{\rho}{r} +... , ~~ \Psi=\frac{\Psi_{-}}{r^{\lambda_{-}}}+\frac{\Psi_{+}}{r^{\lambda_{+}}} + ... ~~ \chi\rightarrow 0, \ \ \ g\rightarrow r^{2}+..., \ \ H\rightarrow 1+...
\label{boundar behavior4D}
\end{eqnarray}
where $\mu$ and $\rho$ are interpreted as the the chemical potential and the charge density of the boundary theory respectively, and 
$\lambda_{\pm}=\frac{3\pm\sqrt{9+4m^2}}{2}$. In this paper we consider a special case with $m^2=-2$ which also implies $\lambda_{\pm}=2,1$. Although $m^2$ is negative but it is
above the Breitenlohner-Freedman (BF) bound $m^2=-9/4$ in four spacetime dimensions. Now some interpretation of the boundary parameters in eq. (\ref{boundar behavior4D}) are in order.
We will interpret the leading falloff $\Psi_{-}$ as the source term and the subleading term $\Psi_{+}\sim O_2$ as the VEV of the dual scalar operator.
With $m^2=-2$, the meaning of $\Psi_{-}$ and $\Psi_{+}$ can also be interchangeable though this scenario is not considered in this paper. Since we want to break the $U(1)$ symmetry
spontaneously, we will set the source term $\Psi_{-}=0$ as the boundary condition.

In the equations of motion (\ref{psieom4D})-(\ref{ttrreinsteineom4D}), we will consider particular forms of $\textrm{G}(\Psi)$ and $\textrm{K}(\Psi)$ :
\begin{equation}
\textrm{G}(\Psi)=\Psi^{2}+\xi \Psi^{\theta},~~~\textrm{K}(\Psi)=\Psi^{2}+\Sigma \Psi^{\gamma}
\label{generalisedterms}
\end{equation}
As in \cite{Dey}, we are mostly interested in examining the phase structure of the boundary superconductors with respect to $\eta$ and $\Sigma$. For this reason we will set the other parameters to
a fixed value, in particular $\xi= 0$ and $\gamma=4$, but we have checked for several examples that a non zero value of $\xi$ and different values of $\gamma$ do not change the results qualitatively.

%%%%%%%%%%%%%%%%%%%%%%%%%%%%%%%%%%%%%%%%%%%%%%%%%%%%%%%%%%%%%%%%%
\begin{figure}[t!]
\begin{minipage}[b]{0.5\linewidth}
\centering
\includegraphics[width=2.7in,height=2.2in]{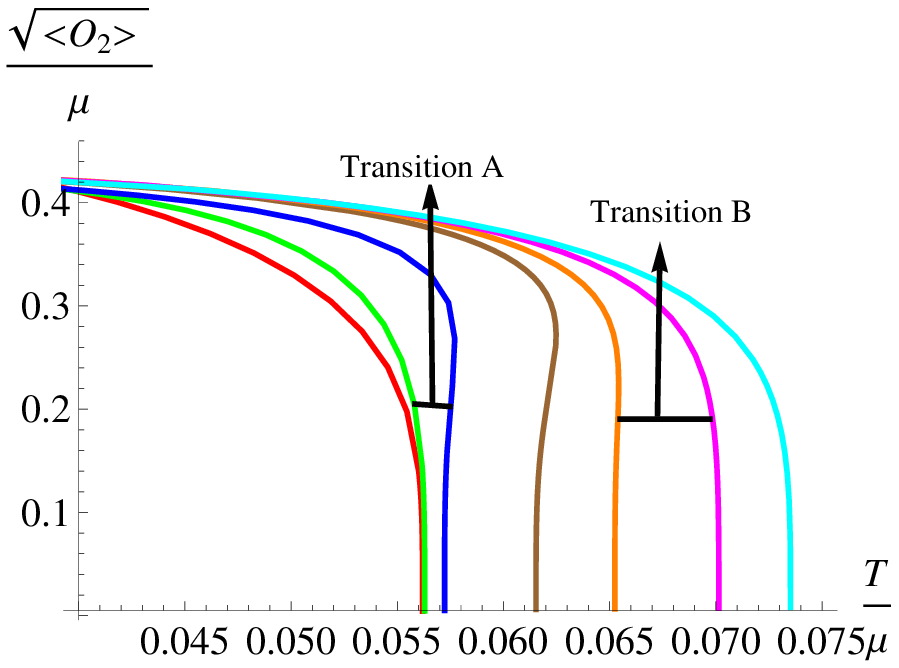}
\caption{Variation of the condensate for different values of $\eta$ with fixed $\Sigma=10$ and $2\kappa^{2}=0.3$ for 4D R-charged black hole background.}
\label{4DO2VsEtaAlpha0.3Sigma10}
\end{minipage}
\hspace{0.4cm}
\begin{minipage}[b]{0.5\linewidth}
\centering
\includegraphics[width=2.7in,height=2.2in]{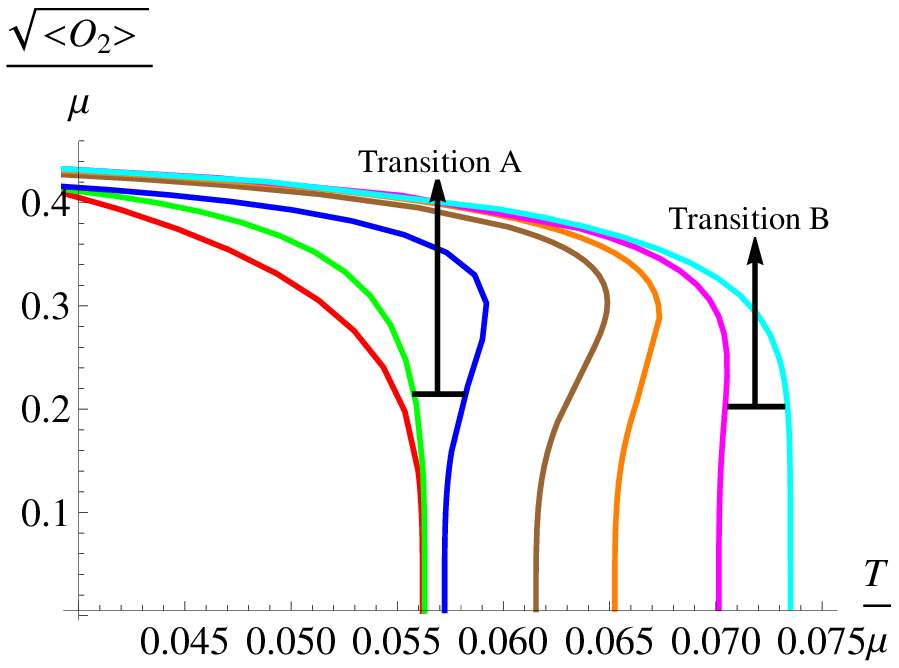}
\caption{Variation of the condensate for different values of $\eta$ with fixed $\Sigma=15$ and $2\kappa^{2}=0.3$ for 4D R-charged black hole background. }
\label{4DO2VsEtaAlpha0.3Sigma15}
\end{minipage}
\end{figure}
%%%%%%%%%%%%%%%%%%%%%%%%%%%%%%%%%%%%%%%%%%%%%%%%%%%%%%%%%%%%%%
%%%%%%%%%%%%%%%%%%%%%%%%%%%%%%%%%%%%%%%%%%%%%%%%%%%%%%%%%%%%%%%%%
\begin{figure}[t!]
\begin{minipage}[b]{0.5\linewidth}
\centering
\includegraphics[width=2.7in,height=2.2in]{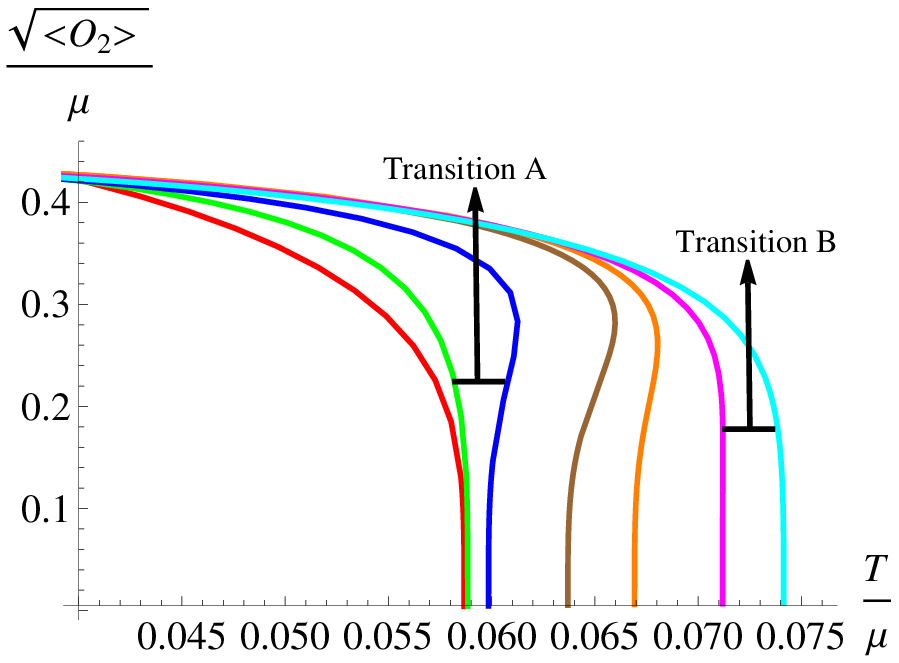}
\caption{Variation of the condensate for different values of $\eta$ with fixed $\Sigma=10$ and $2\kappa^{2}=0$ for 4D R-charged black hole background.}
\label{4DO2VsEtaAlpha0Sigma10}
\end{minipage}
\hspace{0.4cm}
\begin{minipage}[b]{0.5\linewidth}
\centering
\includegraphics[width=2.7in,height=2.2in]{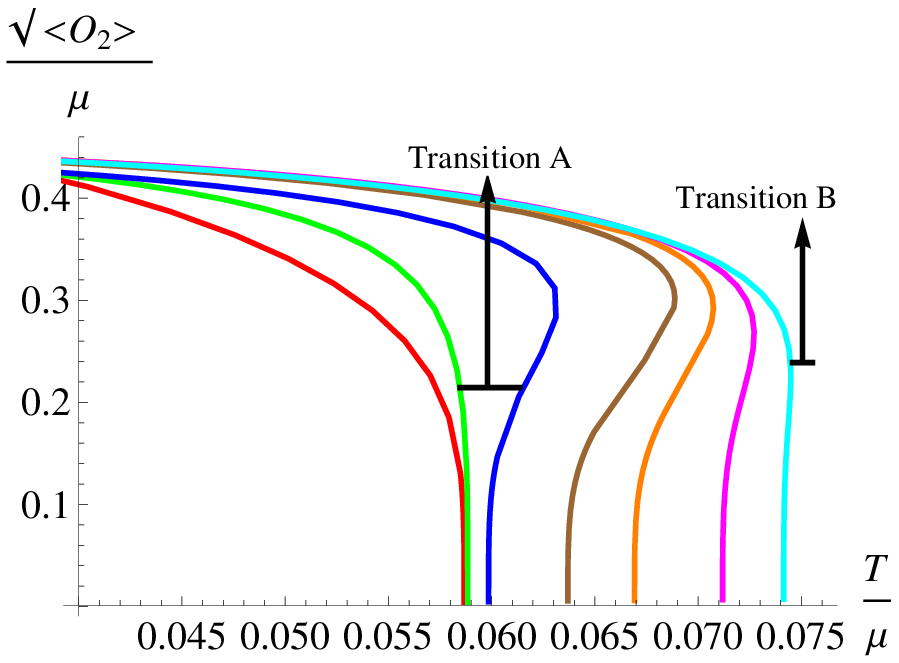}
\caption{Variation of the condensate for different values of $\eta$ with fixed $\Sigma=15$ and $2\kappa^{2}=0$ for 4D R-charged black hole background. }
\label{4DO2VsEtaAlpha0Sigma15}
\end{minipage}
\end{figure}
%%%%%%%%%%%%%%%%%%%%%%%%%%%%%%%%%%%%%%%%%%%%%%%%%%%%%%%%%%%%%%
Now we present numerical results on the VGHS.\footnote{For numerical convenience, we use the $z=r_h/r$ coordinate.}
In figs.(\ref{4DO2VsEtaAlpha0.3Sigma10}) and (\ref{4DO2VsEtaAlpha0.3Sigma15}), we show the plots of the condensate $\sqrt{\langle O_2\rangle}$ with a back reaction
parameter $2\kappa^2 = 0.3$, for various values of $\eta$, with $\Sigma = 10$ and $15$ respectively. In these figures, the red, green, blue, brown, orange, magenta and cyan curves
corresponds to $\eta$=$0.01$, $-0.01$, $-0.1$, $-0.5$, $-1$, $-2$ and $-3$, respectively. One can notice the non zero value of the condensate below a certain critical $T/\mu$ which indicates
the onset of the superconducting phase. Above this $T/\mu$, the system is in the normal phase where the condensate is zero.
We see from figs.(\ref{4DO2VsEtaAlpha0.3Sigma10}) and (\ref{4DO2VsEtaAlpha0.3Sigma15}) that there is an interesting window of first order phase transitions :
as we decrease the higher derivative coupling parameter $\eta$, the transition - which was of second order for
positive values of $\eta$ - changes order, and it remains first order within a range of $\eta$. This range appears to increase with an increase of the value of $\Sigma$.
This is qualitatively indicated in figs.(\ref{4DO2VsEtaAlpha0.3Sigma10}) and (\ref{4DO2VsEtaAlpha0.3Sigma15}), where ``transition A'' refers to the order of the
phase transition changing from second to first, and the reverse for ``transition B.''

In figs.(\ref{4DO2VsEtaAlpha0Sigma10}) and (\ref{4DO2VsEtaAlpha0Sigma15}), we have also shown the condensate calculations for $2\kappa^2=0$, which corresponds to the
probe limit in our model. Analysis of these results indicate that the backreaction parameter for the VGHS in R-charged black hole backgrounds makes the window in $\eta$, with in which the
transition from normal to superconducting phase is first order, narrower compared to the probe limit. This is in stark contrast with the results obtained with the AdS-Schwarzschild black
hole background, where increase in backreaction parameter makes the window in $\eta$ wider compared to the probe limit \cite{Dey}. This is a non-trivial effect of the spin of the brane
configuration.

\begin{figure}[t!]
\begin{minipage}[b]{0.5\linewidth}
\centering
\includegraphics[width=2.7in,height=2.2in]{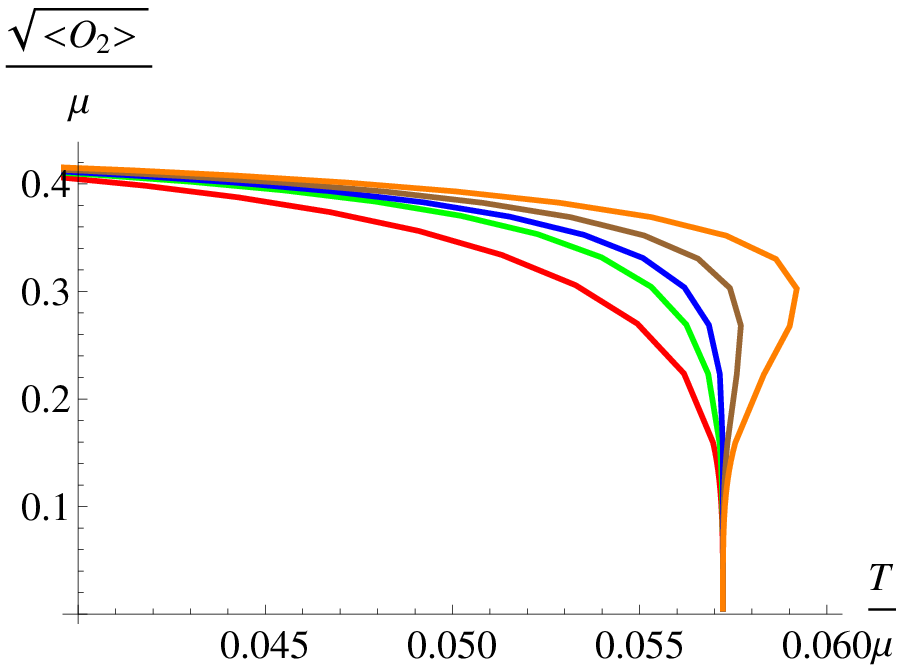}
\caption{Variation of condensate for different values of $\Sigma$ with fixed $\eta=-0.1$ and $2\kappa^{2}=0.3$ for 4D R-charged black hole background.}
\label{4DO2VsSigmaAlpha0.3Eta-0.1}
\end{minipage}
\hspace{0.4cm}
\begin{minipage}[b]{0.5\linewidth}
\centering
\includegraphics[width=2.7in,height=2.2in]{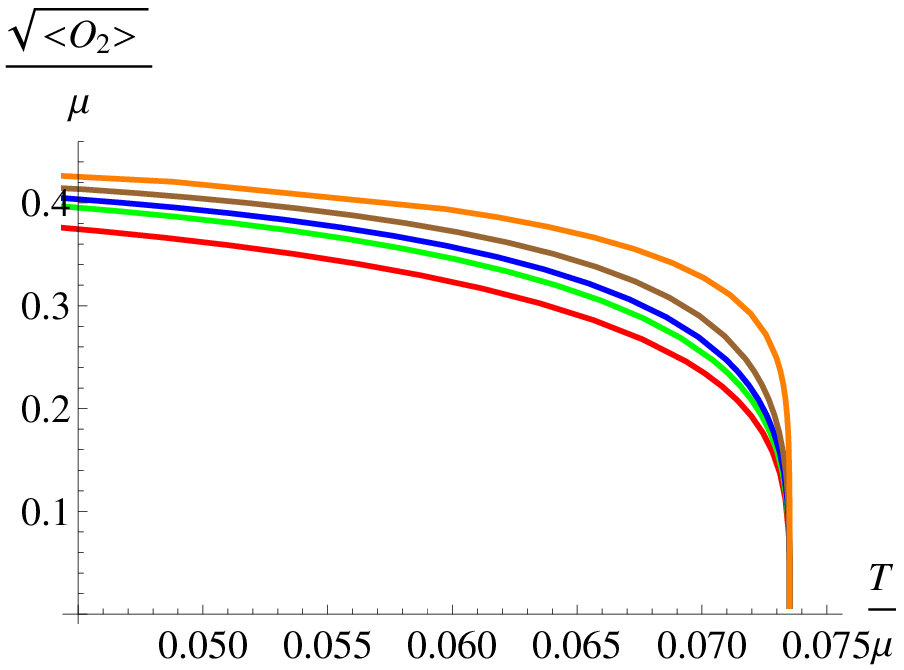}
\caption{Variation of condensate for different values of $\Sigma$ with fixed $\eta=-3$ and $2\kappa^{2}=0.3$ for 4D R-charged black hole background.}
\label{4DO2VsSigmaAlpha0.3Eta-3}
\end{minipage}
\end{figure}

In the same spirit, figs. (\ref{4DO2VsSigmaAlpha0.3Eta-0.1}) and (\ref{4DO2VsSigmaAlpha0.3Eta-3}) show the condensate values for the VGHS, for the same back reaction parameter,
for various values of $\Sigma$, with $\eta$ fixed at $-0.1$ and $-3$
respectively. For both these graphs, the red, green, blue, brown and orange curves corresponds to $\Sigma$=$1$, $5$, $7$, $10$ and $15$, respectively.
We find that for a fixed value of $\eta$, the transition from the normal to the superconductor phase does not have a window (where the transition is of first order), contrary to
the case of fixed $\Sigma$. In the present case, for small negative values of $\eta$, the order of the transition changes
from second to first, as one increases $\Sigma$. This suggest the existence of a lower cutoff in $\Sigma$ ($\Sigma_c$) above which the phase transition from the normal to the
superconducting phase is of first order. For further lower values of $\eta$, in the range of $\Sigma$ considered here, the normal to superconductor transition is always of second order. Qualitatively, this
was the behavior alluded to in the introduction, in fig.(\ref{phasediagram}).
For the sake of comparison with the VGHS in AdS-Schwarzschild black hole backgrounds, we also note that
the value of the cutoff $\Sigma_c$ is larger for our R-charged background.

A word about the magnitude of the critical $T/\mu$ is in order. Normally, higher backreaction parameter makes the critical $T/\mu$ smaller, which generally implies that the backreaction makes the
scalar condensation harder to form. This is also the case here.
In a similar manner, the critical value of $T/\mu$ also decreases for higher values of $\eta$ but, on the other hand, does not depends on $\Sigma$.
This is expected from a physical ground in eq. (\ref{generalisedterms}), since at the phase transition point $\Psi$ is negligible and therefore $\Sigma$ which comes with higher powers of $\Psi$
does not have any effect on critical $T/\mu$. We mention here the overall behavior of critical $T/\mu$ in R-charged black hole backgrounds for different value of $\kappa$
and $\eta$, is the same as in AdS-Schwarzschild black hole background but with higher magnitude. This indicates that the scalar field instability is easier to form in a VGHS for
spinning brane configurations.
%%%%%%%%%%%%%%%%%%%%%%%%%%%%%%%%%%%%%%%%%%%%%%%%%%%%%%%%%%%%%%%%%
%%%%%%%%%%%%%%%%%%%%%%%%%%%%%%%%%%%%%%%%%%%%%%%%%%%%%%%%%%%%%%

To check the validity of this result,
it is worthwhile to understand the behavior of the free energy in these cases. To highlight the essential physics, it is enough to consider the probe limit, with $\kappa^2=0$ and compute the Gibbs free
energy of the boundary thermal state by identifying the latter with the bulk on-shell action. As usual, one has to add a boundary counter term to the on-shell action, and calculate the renormalized
free energy. Following this procedure, we find that the difference of free energy between the normal and the superconducting phase is given by the expression
\begin{eqnarray}
\Delta\Omega &=& \Omega_{Superconductor}-\Omega_{Normal}\nonumber\\
&=& -{\mu \rho \over 4}+{1 \over 2}\int_0^1 \mathrm{d}z {\Phi(z)^2 \Psi(z)^2\over z^4 g(z)}
+{\eta \over 2 }\int_0^1 \mathrm{d}z z^4 g(z) \Phi'(z)^2 \Psi'(z)^2 \nonumber\\
&-&{\eta \over 2}\int_0^1 \mathrm{d}z \frac{\Phi(z)^2 \Phi'(z)^2\Psi(z)^2}{g(z)}\Big(2+3\Sigma \Psi(z)^2\Big)
+{\mu^2\over 4}\nonumber
 \end{eqnarray}
%%%%%%%%%%%%%%%%%%%%%%%%%%%%%%%%%%%%%%%%%%%%%%%%%%%%%%%%%%%%%%%%%
\begin{figure}[t!]
\begin{minipage}[b]{0.5\linewidth}
\centering
\includegraphics[width=2.7in,height=2.2in]{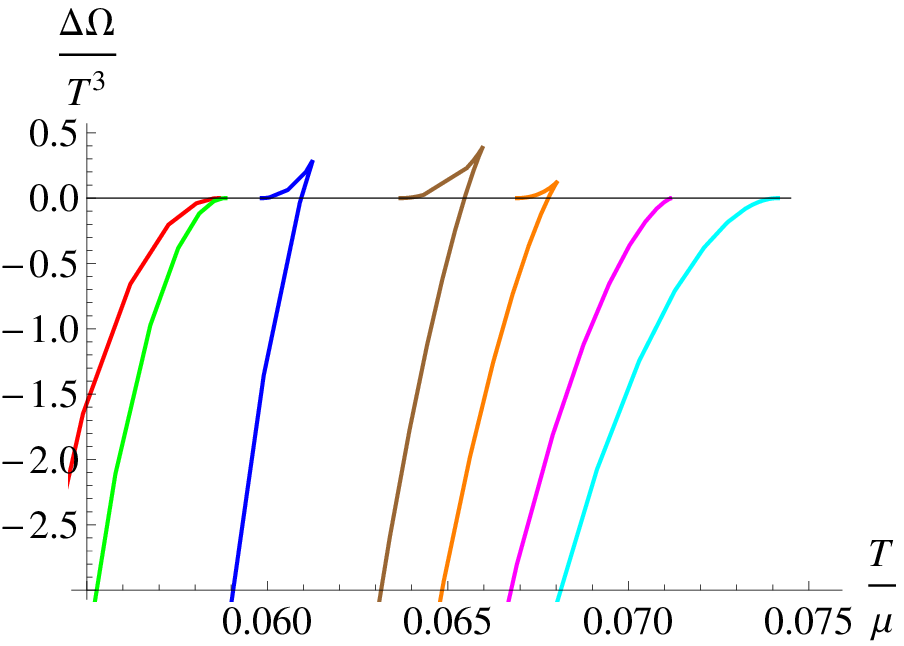}
\caption{Difference in free energy between the superconducting and normal phase in 4D R charged background for fixed $\Sigma=10$ and $2\kappa^{2}=0$ for different values of
$\eta$. Here we have used the same color coding as in fig. (\ref{4DO2VsEtaAlpha0.3Sigma10}).}
\label{4DFreeEnergyVsEtaAlpha0Sigma10}
\end{minipage}
\hspace{0.4cm}
\begin{minipage}[b]{0.5\linewidth}
\centering
\includegraphics[width=2.7in,height=2.2in]{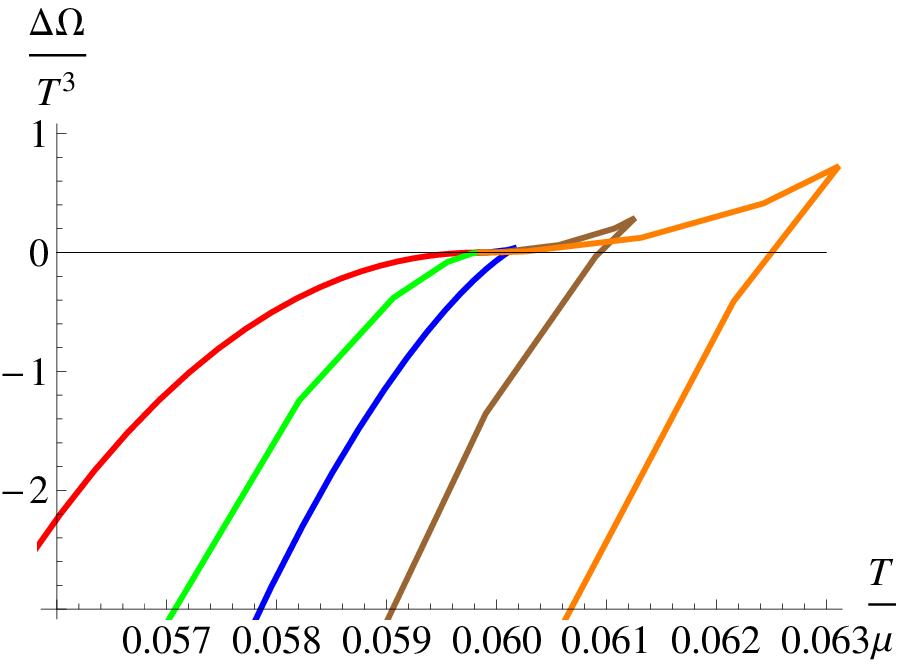}
\caption{Difference in free energy between the superconducting and normal phase in 4D R charged background for fixed $\eta=-0.1$ and $2\kappa^{2}=0$ for different values of
$\Sigma$. Here we have used the same color coding as in fig. (\ref{4DO2VsSigmaAlpha0.3Eta-0.1}).}
\label{4DFreeEnergyVsSigmaAlpha0Eta-0.1}
\end{minipage}
\end{figure}
%%%%%%%%%%%%%%%%%%%%%%%%%%%%%%%%%%%%%%%%%%%%%%%%%%%%%%%%%%%%%%
This difference of the free energy is plotted in figs.(\ref{4DFreeEnergyVsEtaAlpha0Sigma10}) and(\ref{4DFreeEnergyVsSigmaAlpha0Eta-0.1}) which use the same colour coding as
fig.(\ref{4DO2VsEtaAlpha0.3Sigma10}) and fig.(\ref{4DO2VsSigmaAlpha0.3Eta-3}), respectively. The result is exactly as we expect. For fixed $\Sigma$, we find a window
of first order phase transitions, which is absent in the analysis with fixed $\eta$. A similar result was obtained in \cite{Dey} for VGHS in the AdS-Schwarzschild background.
%%%%%%%%%%%%%%%%%%%%%%%%%%%%%%%%%%%%%%%%%%%%%%%%%%%%%%%%%%%%%%%%%
\begin{figure}[t!]
\begin{minipage}[b]{0.5\linewidth}
\centering
\includegraphics[width=2.7in,height=2.2in]{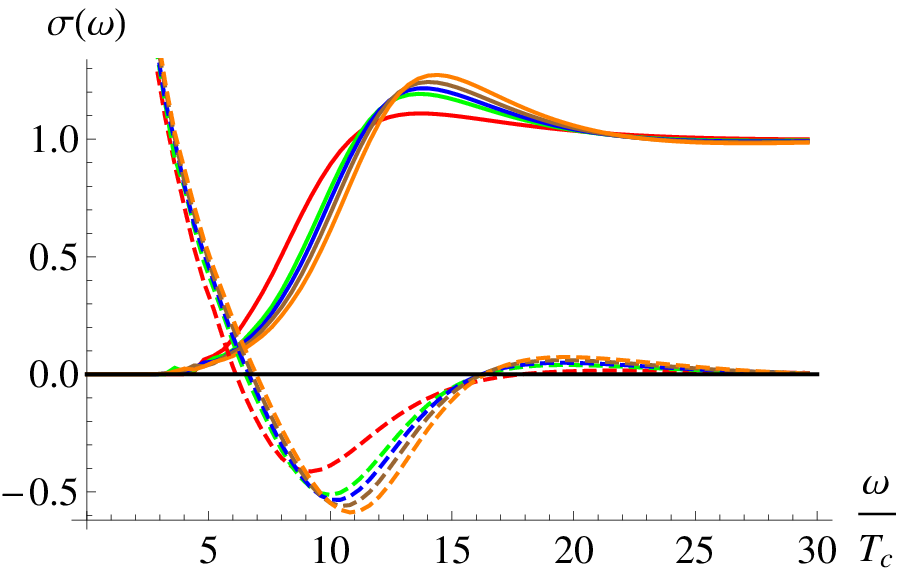}
\caption{Real (solid lines) and imaginary (dotted lines) part of conductivity for different values of $\Sigma$ with fixed $\eta=-0.1$ and $2\kappa^{2}=0$ for 4D R-charged black hole background.}
\label{4DCondVsSigmaAlpha0Eta-0.1}
\end{minipage}
\hspace{0.4cm}
\begin{minipage}[b]{0.5\linewidth}
\centering
\includegraphics[width=2.7in,height=2.2in]{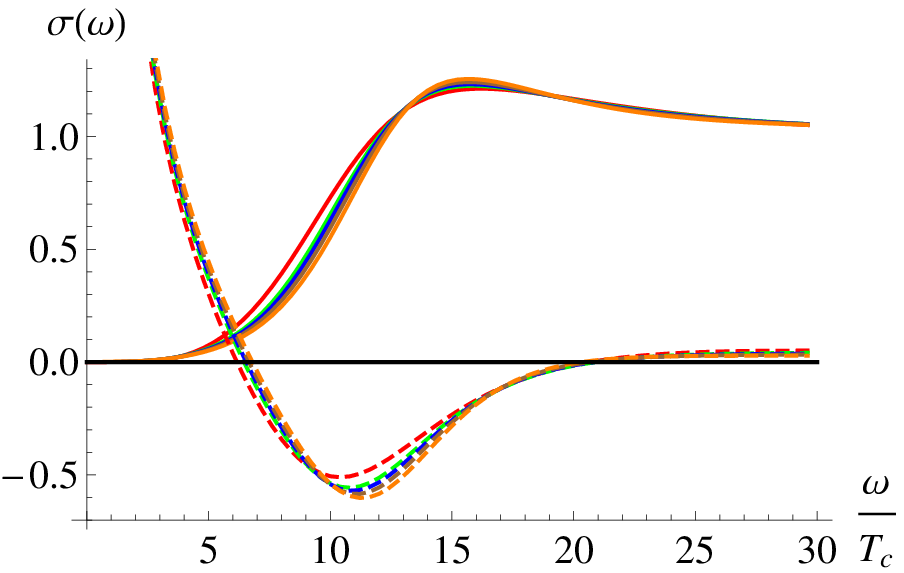}
\caption{Real (solid lines) and imaginary (dotted lines) part of conductivity for different values of $\Sigma$ with fixed $\eta=-0.1$ and $2\kappa^{2}=0.3$ for 4D R-charged black hole background.}
\label{4DCondVsSigmaAlpha0.3Eta-0.1}
\end{minipage}
\end{figure}
%%%%%%%%%%%%%%%%%%%%%%%%%%%%%%%%%%%%%%%%%%%%%%%%%%%%%%%%%%%%%%

We now proceed to calculate the optical conductivity of our boundary superconducting system. We work with vector type perturbations in the metric and in the gauge field, with
$g_{xt} \neq 0$, $g_{xy} \neq 0$ and $A_{x} \neq 0$. The computation is standard: we assume the spatial and time dependence of the perturbations to be of the form $e^{iky - i\omega t}$,
and work at the linearized level. In this perturbation there are four independent equations. However in the limit $k \to 0$, which is appropriate to compute the optical conductivity,
two of these independent equations - namely the $xt$ and the $xy$ components of the Einstein equations - decouple. After rearranging the other two equations, we find a
second order differential equation for $A_{x}$, which is given by 
\begin{eqnarray}
&A_{x}''&\left(1+\frac{2\eta g \psi'^2}{H^2} \right)+ A_{x}'\left(\frac{g'}{g}+\frac{H'}{H}-\frac{\chi '}{2} \right)+ \frac{\eta \psi'^2 A_{x}'}{H^2}\left(4g' -g \chi' +
\frac{4 g \psi''}{\psi'} -\frac{2gH'}{H} \right) \nonumber \\  &+&  2\kappa^{2} \eta ^2 e^{\chi} A_{x} \left(-\frac{2e^{2\chi}   \phi^4 \phi'^2 \text{K}(\Psi)^2}{g^3} +
\frac{4 e^{\chi}\phi^2 \phi'^2 \psi'^2 \text{K}(\Psi)}{gH} - \frac{2g \phi'^2 \psi'^4}{H^2}\right) \nonumber \\  &+&
A_{x} \left(\frac{e^{\chi}H \omega^2}{g^2}-\frac{2\text{G}(\Psi)}{gH}-\frac{2\kappa^{2} e^{\chi}H^{2}\phi'^2}{2g} \right) +  \eta e^{\chi}\textrm{K}(\Psi) A_{x}
\biggl(-\frac{2e^{\chi} \omega^2\phi^2}{g^3} \nonumber \\  &-& \frac{2\phi \phi' \textrm{K}(\Psi)'}{g H \textrm{K}(\Psi)}-\frac{2\phi'^2}{gH}+
\frac{4\kappa^{2} e^{\chi}H \phi^2 \phi'^2}{g^2}-\frac{\phi \phi' \chi'}{gH}- \frac{4\kappa^{2}\phi'^2 \psi'^2}{\text{K}(\Psi)}-\frac{2\phi \phi''}{gH} \biggr)=0 \nonumber \\
\label{Axeom4D}
\end{eqnarray}
where we have again suppressed the $r$-dependence. In order to solve this equation we need to apply appropriate boundary conditions. At the horizon, we impose an infalling
boundary condition $A_{x} \ \alpha \  g(r)^{-i \omega/{4 \pi T_{h}}}$. At the asymptotic boundary, $A_{x}$ behaves as
\begin{equation}
A_{x}=A_{x}^{(0)}+\frac{A_{x}^{(1)}}{r}+...
\end{equation}
Using the AdS/CFT prescription, one can identify the leading term $A_{x}^{(0)}$ and the subleading term $A_{x}^{(1)}$ as the dual source and
the expectation value of boundary current, respectively, and the expression for the conductivity by calculating current-current correlator is given by \cite{Son}
$$\sigma(\omega)=-\frac{i A_{x}^{(1)}}{\omega A_{x}^{(0)}}$$
The results are shown in figs.(\ref{4DCondVsSigmaAlpha0Eta-0.1}) and (\ref{4DCondVsSigmaAlpha0.3Eta-0.1}), where we have fixed $\eta = -0.1$ and the red, green, blue, brown and orange curves
correspond to $\Sigma$=$1$, $5$, $7$, $10$ and $15$, respectively. While fig.(\ref{4DCondVsSigmaAlpha0Eta-0.1}) corresponds to the probe limit $\kappa^2 = 0$, fig.(\ref{4DCondVsSigmaAlpha0.3Eta-0.1})
is for a non-zero value of the back reaction, $2\kappa^2 = 0.3$. \footnote{Here the temperature is measured in units of $\rho$ and we have chosen $T=0.2 T_c$.}
From these figures, we can see the signature of the pole in the imaginary part of the conductivity at $\omega=0$. It implies that, using Kramers-Kronig relations which relate
the real and imaginary part of the conductivity, the real part of the conductivity has a delta function at $\omega=0$. However, this delta function is not visible in the numerical calculations
in fig.(\ref{4DCondVsSigmaAlpha0Eta-0.1}) and (\ref{4DCondVsSigmaAlpha0.3Eta-0.1}) due to its infinitesimal width.  Another important observation from these figures is the magnitude
of gap frequency to the critical temperature, $\omega_g/T_c \sim 10$, where $\omega_g$ is defined as the frequency at which the imaginary part of the conductivity is minimum.
Interestingly, this ratio is relatively small compared to the VGHS in AdS-Schwarzschild black hole background where $\omega_g/T_c$ was found to be nearly $20$ \cite{Dey}.
This indicates that the boundary superconductor in an AdS-Schwarzschild black hole background is more strongly coupled than its R-charged cousin. 
For different value of $\eta$, the results for the conductivity are qualitatively similar.

\section{5-D R-charged black hole backgrounds}

For 5-D R-charged backgrounds, the procedure for constructing a VGHS is entirely similar to what has been discussed in the previous section.
We will relegate the details of the computation here to Appendix A, and simply present numerical results. Here we have considered $m^2=-15/4$, which is again above the
Breitenlohner-Freedman bound $m_{BF}^2=-4$ for the five dimensional AdS background.\footnote{For the interpretation of various physical quantities, see Appendix A.}

In fig.(\ref{5DO2VsEtaAlpha0.3Sigma10}), we have plotted the condensate as a function of the temperature for 5-D R-charge backgrounds
with a fixed values of $\Sigma = 10$ and the back reaction parameter $2\kappa^2=0.3$, for different values of the higher derivative coupling parameter $\eta$. Here, the
red, green, blue, brown, orange, magenta and cyan curves corresponds to $\eta$ = $0.01$, $-0.01$, $-0.1$, $-0.5$, $-1$, $-2$ and $-3$, respectively. We again get a phase 
transition from normal to superconducting phase below a critical $T/\mu$ and find a window in $\eta$ for the first order phase transitions. 
However, this window is relatively larger compared to the 4D R-charged case (fig. (\ref{4DO2VsEtaAlpha0.3Sigma10})). The same scenario (not presented here)
is observed in the probe limit also. 

%%%%%%%%%%%%%%%%%%%%%%%%%%%%%%%%%%%%%%%%%%%%%%%%%%%%%%%%%%%%%%%%%
\begin{figure}[t!]
\begin{minipage}[b]{0.5\linewidth}
\centering
\includegraphics[width=2.7in,height=2.2in]{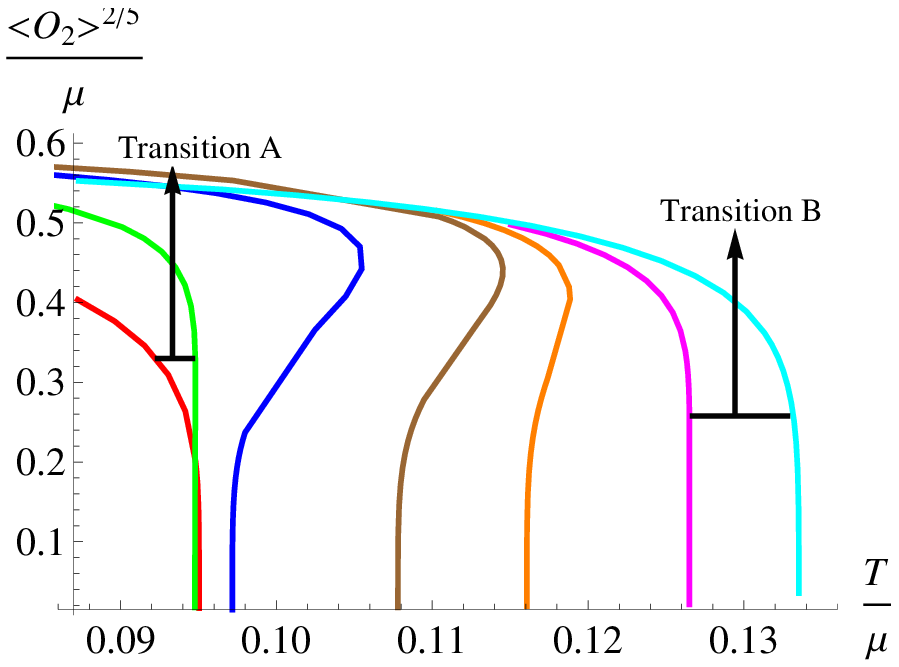}
\caption{Variation of the condensate for different values of $\eta$ with fixed $\Sigma=10$ and $2\kappa^{2}=0.3$ for 5D R-charged black hole background.}
\label{5DO2VsEtaAlpha0.3Sigma10}
\end{minipage}
\hspace{0.4cm}
\begin{minipage}[b]{0.5\linewidth}
\centering
\includegraphics[width=2.7in,height=2.2in]{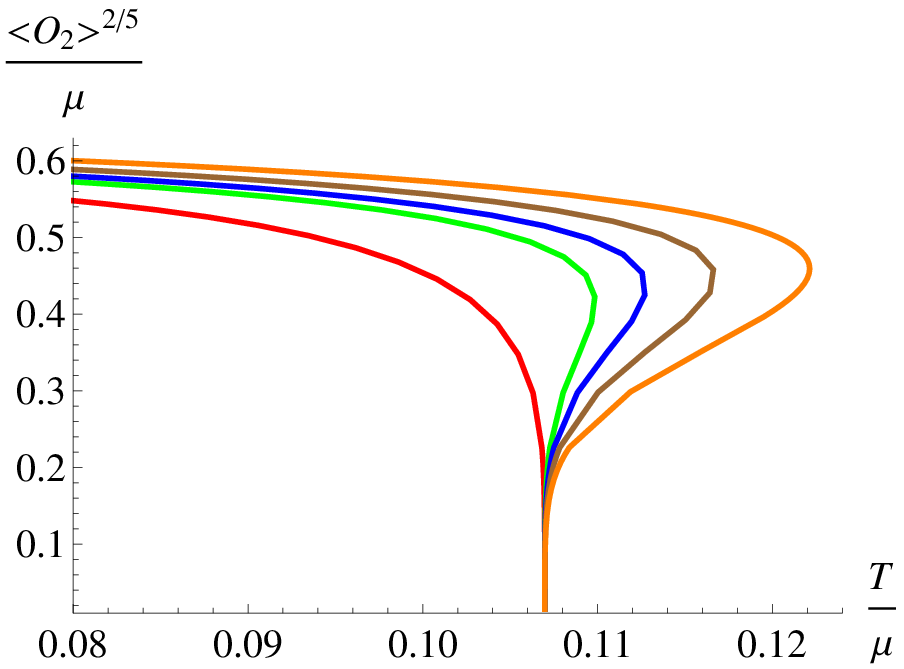}
\caption{Variation of the condensate for different values of $\Sigma$ with fixed $\eta=-0.1$ and $2\kappa^{2}=0$ for 5D R-charged black hole background.}
\label{5DO2VsSigmaAlpha0Eta-0.1}
\end{minipage}
\end{figure}
%%%%%%%%%%%%%%%%%%%%%%%%%%%%%%%%%%%%%%%%%%%%%%%%%%%%%%%%%%%%%%
%%%%%%%%%%%%%%%%%%%%%%%%%%%%%%%%%%%%%%%%%%%%%%%%%%%%%%%%%%%%%%%%%
\begin{figure}[h!]
\begin{minipage}[b]{0.5\linewidth}
\centering
\includegraphics[width=2.7in,height=2.2in]{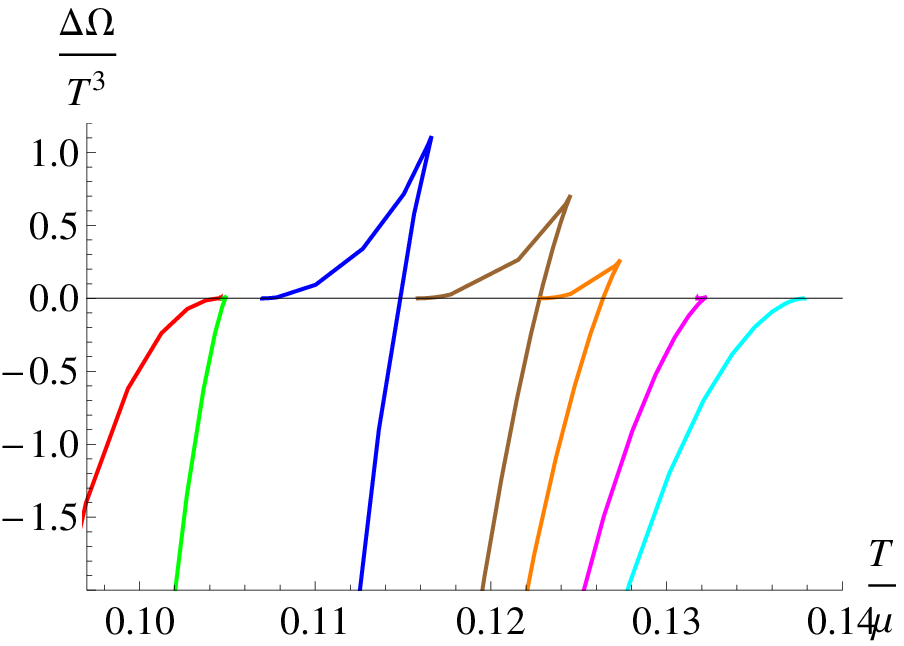}
\caption{Difference in free energy between the superconducting and normal phase in 5-D R charged background for fixed $\Sigma=10$ and $2\kappa^{2}=0$ for different values of $\eta$.}
\label{5DFreeEnergyVsEtaAlpha0Sigma10}
\end{minipage}
\hspace{0.4cm}
\begin{minipage}[b]{0.5\linewidth}
\centering
\includegraphics[width=2.7in,height=2.2in]{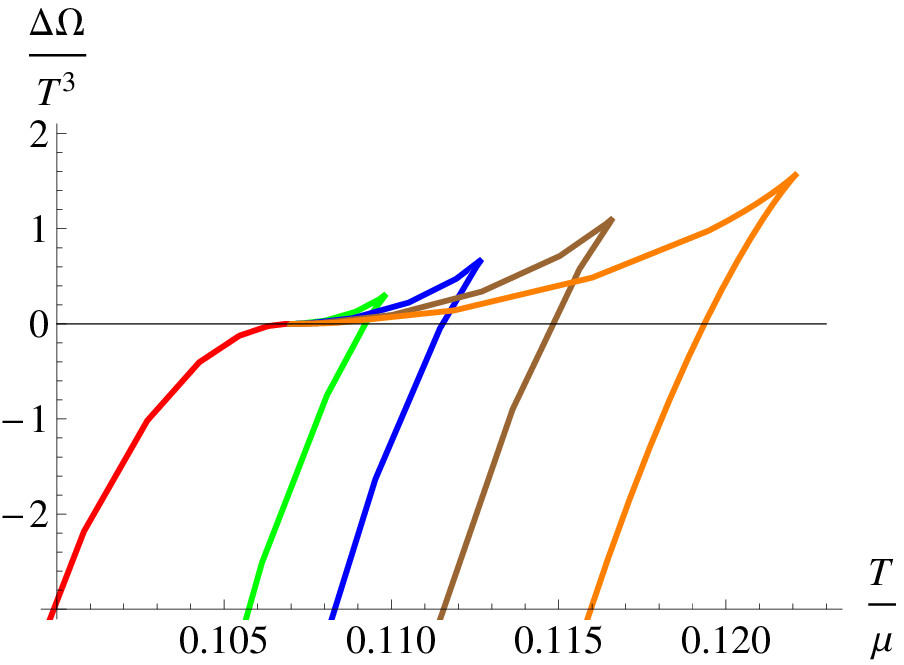}
\caption{Difference in free energy between the superconducting and normal phase in 5-D R charged background for fixed $\eta=-0.1$ and $2\kappa^{2}=0$ for different values of $\Sigma$.}
\label{5DFreeEnergyVsSigmaAlpha0Eta-0.1}
\end{minipage}
\end{figure}
%%%%%%%%%%%%%%%%%%%%%%%%%%%%%%%%%%%%%%%%%%%%%%%%%%%%%%%%%%%%%%
In the probe limit, the condensate is plotted for different values of $\Sigma$ in fig.(\ref{5DO2VsSigmaAlpha0Eta-0.1}), where
the red, green, blue, brown and orange curves correspond to $\Sigma$=$1$, $5$, $7$, $10$ and $15$, respectively. We see that the results are qualitatively similar to our computation 
in the 4-D background with a lower cutoff parameter $\Sigma_c$, above which the transition is always of first order.

These results were checked with the corresponding free energy calculations, which are presented in figs.(\ref{5DFreeEnergyVsEtaAlpha0Sigma10}) and 
(\ref{5DFreeEnergyVsSigmaAlpha0Eta-0.1}).
In both these figures, we have chosen the back reaction to be zero, for illustration.  In the former case, we get a window of first order phase transitions which is absent in the 
latter thereby justifying the results we obtained by analysing the condensate.

\section{\textbf{HEE for very general holographic superconductors}}

In this section, we will compute the holographic entanglement entropy for very general holographic superconductors. For the sake of completeness, we will first recapitulate
a few known facts. As mentioned in the introduction, entanglement entropy is a measure of the correlation between two subsystems $\mathcal{A}$ and $\mathcal{B}$ of a given quantum system.
Specifically, the entanglement entropy of subsystem $\mathcal{A}$ is given by,
\begin{equation}
 S_{\mathcal{A}}=-\mathrm{Tr}_{\mathcal{A}} (\rho_{\mathcal{A}} \ln
\rho_{\mathcal{A}}) ~.
 \label{EE1}
\end{equation}
where $\rho_{\mathcal{A}}$ is the reduced density matrix of $\mathcal{A}$, calculated by taking the trace over the degrees of  freedom of $\mathcal{B}$, i.e,
$\rho_{\mathcal{A}}=\mathrm{Tr}_{\mathcal{B}}(\rho)$, $\rho$ being the density matrix of the full quantum system. In a holographic setup, the Ryu-Takayanagi proposal states that
the HEE of the subsystem $\mathcal{A}$ living on the boundary of a $(d+1)$ dimensional AdS space is given by,
\begin{equation}
S_{\mathcal{A}} = {\mbox{Area}(\gamma_\mathcal{A})\over 4 G_N} ~
\label{HEE}
\end{equation}
where $G_N$ is the gravitational constant in $(d+1)$ dimension and $\gamma_\mathcal{A}$ is the $(d-1)$ dimensional
minimal-area hypersuface which extends into the bulk and has the same boundary $\partial\mathcal{A}$ of the subsystem $\mathcal{A}$.

Several computations of the HEE has been performed using the Ryu-Takayanagi prescription, and they are in good agreement with CFT
results. For example, using standard techniques, one can compute the EE for a subsystem of length $l$ in a 2D CFT, which is given by $S_{\mathcal{A}} = \frac{c}{3} \ln {l \over
\epsilon}$ where $c$ is the central charge of the CFT and $\epsilon$ is an UV cut-off \cite{Cardy}. Instead, using AdS$_3$/CFT$_2$, if we apply
the Ryu-Takayanagi formula for the HEE, we get the same result, with $c=3R/2G_{N}^{(3)}$, where $R$ is radius of curvature of
AdS$_3$ and $G_{N}^{(3)}$ is the three dimensional gravitational constant.

Now we calculate the entanglement entropy of the VGHS and study the effect of the higher derivative coupling term $\eta$ and the model parameter $\Sigma$
on its HEE.
First, as a warm up exercise, we calculate the HEE for the VGHS in an AdS-Schwarzschild background. Since the necessary
formulas were worked out in \cite{Dey}, we do not show them here, but for completeness reproduce them in Appendix B.
The strategy of the computation is standard. Having solved the coupled equations in the bulk and thus having found the gravity solution both in the superconducting phase
as well as in the normal phase, we use the Ryu-Takayanagi prescription to determine the HEE for both the normal and the superconducting phases. For this we consider our subsystem $\mathcal{A}$
to be a straight strip residing on the boundary. The domain $-\frac{l}{2}\leq x \leq \frac{l}{2}$ and $0\leq y \leq L_0$, defines the strip geometry on the boundary, where $l$ is the size of
region $\mathcal{A}$ and $L_0$ is a regulator which we can later set to infinity. Now we parameterize the minimal surface $\gamma_\mathcal{A}$, which extends in the bulk,
by $x=x(z)$ and calculate the area of this hypersurface using the metric of eq.(\ref{metric4d}). This is given as
\begin{equation}\label{Area}
\mbox{Area}(\gamma_\mathcal{A})=L_0\int_{-l/2}^{l/2}\frac{dx}{z^2}\sqrt{1+\frac{z'(x)^2}{f(z)}}~.
\end{equation}
Minimization of this area functional yields,
\begin{equation}\label{zstar}
\frac{1}{z^2}{1 \over \sqrt{1+\frac{z'(x)^2}{f(z)}}}=\frac{1}{z_*^2}
\end{equation}
where $z_{*}$ is the turning point of the minimal area such that $z'(x)|_{z=z_{*}}=0$. Finally, one can obtain the entanglement entropy \cite{Ryu} as
\begin{equation}\label{EE}
S={\mbox{Area}(\gamma_\mathcal{A})\over 4 G_4}={2L_0\over 4G_4}\int_{\epsilon}^{z_{*}}dz\frac{z_{*}^2}{z^2}
\frac{1}{\sqrt{(z_{*}^4-z^4)f(z)}}={2L_0\over 4G_4}(s+\frac{1}{\epsilon})~,
\end{equation}
with
\begin{equation}\label{length}
\frac{l}{2}=\int_{\epsilon}^{z_{*}}dz\frac{z^2}{\sqrt{(z_{*}^4-z^4)f(z)}}
\end{equation}
where in eq. (\ref{EE}) the first term $s$ is the finite part of entanglement entropy. We also see that the second term in this equation diverges as $\epsilon\rightarrow0$
and $z=\epsilon$ defines the UV cutoff. Since the finite part $s$ does not depend on any cutoff, it is the quantity which is physically important.
So in the rest of our calculations, we will only deal with the finite part of the entanglement entropy.

For comparison, We will first show the results for the condensate for AdS-Schwarzschild black hole background, using the formulas presented in appendix B. 
Fig.(\ref{O2-TSigma5}) shows how the condensate grows as one
decreases the temperature below the critical value of $T/\mu$ for $\Sigma=5$, where the red, green, blue, brown, orange,
pink and cyan curves correspond to $\eta$ = 0.01, -0.01, -0.1, -0.5, -1, -3, and -5 respectively.
Fig.(\ref{O2-Teta-0.1}) shows the behavior of condensate as a function of temperature for $\eta=-0.1$ where the red, green, blue,
brown and orange curves correspond to $\Sigma$ = 0, 1, 3, 5 and 7 respectively. For detail on the analysis of this model, see \cite{Dey}.

\begin{figure}[h!]
\begin{minipage}[b]{0.5\linewidth}
\centering
\includegraphics[width=2.7in,height=2.3in]{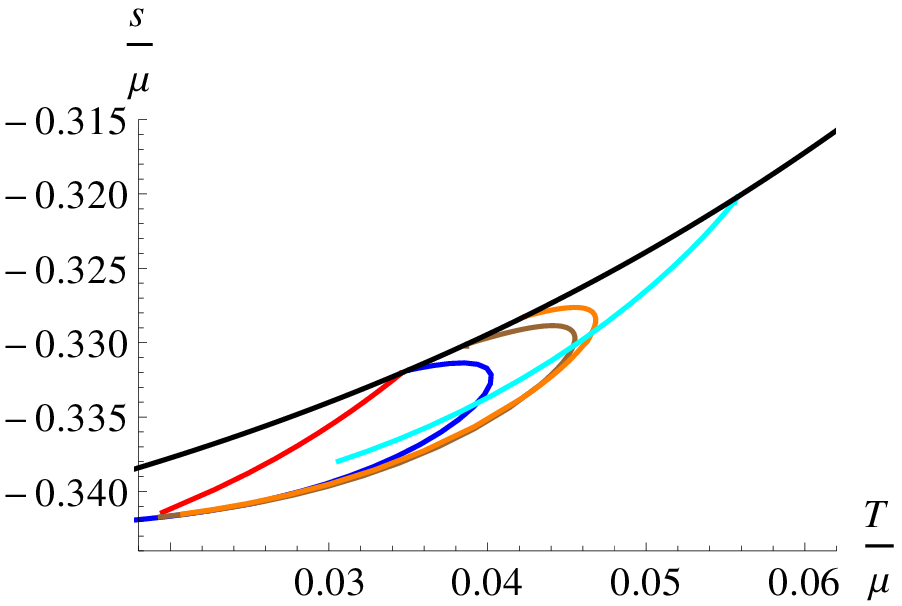}
\caption{HEE for fixed ${l\over 2}\mu=1$, $\Sigma=5$ and $2\kappa^{2}=0.5$ for different values of $\eta$.}
\label{s-TSigma5}
\end{minipage}
\hspace{0.4cm}
\begin{minipage}[b]{0.5\linewidth}
\centering
\includegraphics[width=2.7in,height=2.3in]{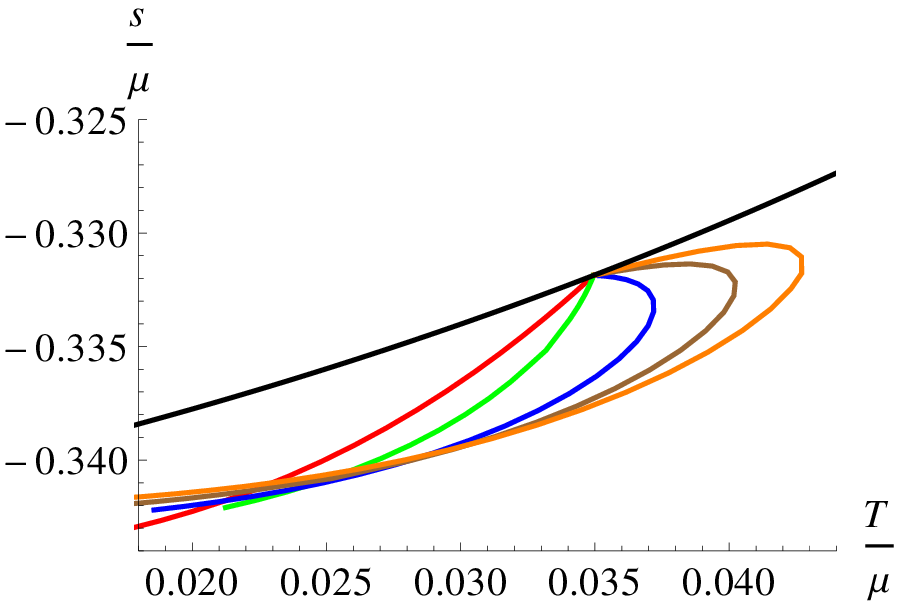}
\caption{HEE for fixed ${l\over 2}\mu=1$, $\eta=-0.1$, $2\kappa^{2}=0.5$ for different values of $\Sigma$.}
\label{s-Teta-0.1}
\end{minipage}
\end{figure}

Keeping in mind that the dimensionless quantities here are ${T\over \mu}$, ${s\over \mu}$ and $l \mu$, we first examine how the HEE changes when we vary the temperature, while keeping
the strip width fixed. We set ${l\over 2}\mu=1$, $\Sigma=5$, $2\kappa^2=0.5$ and consider different values of $\eta$. The results are shown in Fig.(\ref{s-TSigma5}) where the
same color coding as fig.(\ref{O2-TSigma5}) has been used, and the solid
black curve denotes the HEE for the normal phase. For $\eta=0.01$, there is a discontinuity in the slope of $s$ at the critical value of
${T\over \mu}$, which indicates a second order phase transition from normal to superconducting phase \cite{Johnson}\cite{Kuang_1}.
%%%%%%%%%%%%%%%%%%%%%%%%%%%%%%%%%%%%%%%%%%
\begin{figure}[t!]
\begin{minipage}[b]{0.5\linewidth}
\centering
\includegraphics[width=2.7in,height=2.3in]{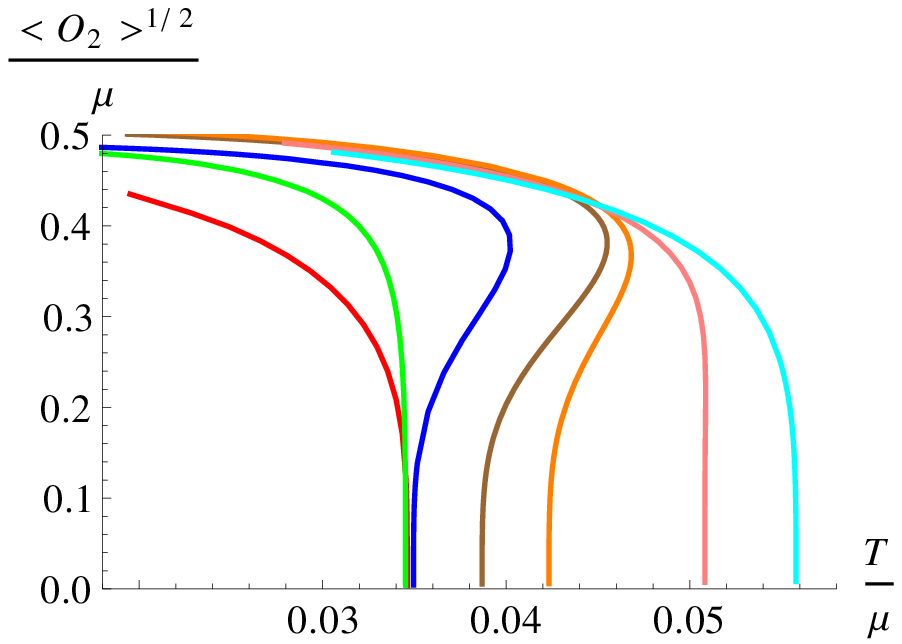}
\caption{Variation of the condensate for different values of $\eta$ with fixed $\Sigma=5$ and $2\kappa^{2}=0.5$ for 4D AdS-Schwarzschild black hole backgrounds.}
\label{O2-TSigma5}
\end{minipage}
\hspace{0.4cm}
\begin{minipage}[b]{0.5\linewidth}
\centering
\includegraphics[width=2.7in,height=2.3in]{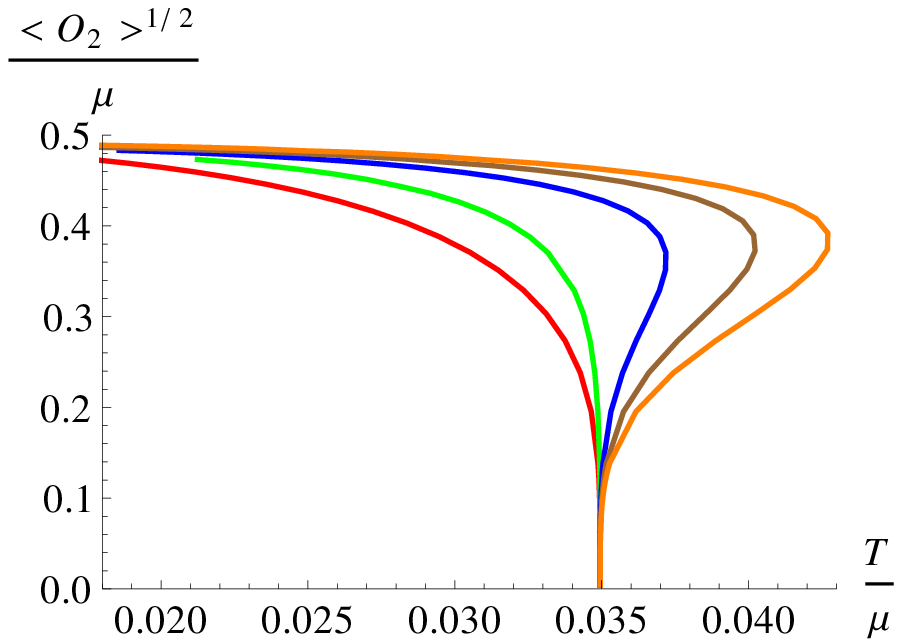}
\caption{Variation of the condensate for different values of $\Sigma$ with fixed $\eta=-0.1$ and $2\kappa^{2}=0.5$ for 4D AdS-Schwarzschild black hole backgrounds.}
\label{O2-Teta-0.1}
\end{minipage}
\end{figure}
%%%%%%%%%%%%%%%%%%%%%%%%%%%%%%%%%%%%%%%%%%
%%%%%%%%%%%%%%%%%%%%%%%%%%%%%%%%%%%%%%%%%%
%%%%%%%%%%%%%%%%%%%%%%%%%%%%%%%%%%%%%%%%%%

As we decrease the value of $\eta$ from 0.01 we see that $s$ becomes multivalued near the critical value of $T/\mu$ and that there is a discontinuous jump in the value of
$s$ at the transition point, which indicates a first order phase transition \cite{Johnson}. If we continue to decrease the value of $\eta$ the transition again
becomes of second order. Indeed, from Fig.(\ref{s-TSigma5}) we see that $\eta=-0.1,-0.5,-1$ give first order phase transitions, while $\eta=-5$ gives second order phase transition.
Thus, like the free energy calculations, the HEE in the VGHS also tells us that for a fixed value of $\Sigma$ and $\kappa$ there exists a window in $\eta$ where the transition from the
normal phase to the superconducting phase is of first order, but outside this window the transition is of second order. This agrees perfectly with our result on condensate as a function
of temperature which is shown in Fig.(\ref{O2-TSigma5}). We notice that for a fixed value of the strip width, the superconducting
solution always has lower entanglement entropy than the normal solution, consistent with our previous discussion. However for the VGHS in R-charged black hole backgrounds, 
which we momentarily turn to, we will find that this result can change, namely the HEE in the superconducting phase can be higher than that in the normal phase. 

Now we will analyze the HEE as a function of $\Sigma$, for fixed $\eta$. This is shown in Fig.(\ref{s-Teta-0.1}), where we have set ${l\over 2}\mu=1$ and $2\kappa^2=0.5$.
In fig.(\ref{s-Teta-0.1}), the same color coding as in fig.(\ref{O2-Teta-0.1}) has been used.
We see that for $\Sigma=0$ and $1$, the transition is second order, but if we increase the value of $\Sigma$, there is a discontinuous jump in $s$ after a certain value of $\Sigma$, indicating
a first order transition. This implies that, for a fixed value of $\eta$ and $\kappa$, there exists a lower cut-off $\Sigma_c$ above which the phase transition is always of first order.
This again agrees with our earlier findings. We have checked for a number of cases that as $\eta$ becomes more and more negative, the cut-off value  $\Sigma_c$ increases.

We record a further observation regarding the magnitudes of the entanglement entropy $s$. At a fixed temperature as we increase $\Sigma$, $s$ first decreases but if
we continue to increase the value of $\Sigma$, at a certain point $s$ starts to increase. However, this behavior depends on temperature. For example, at ${T\over \mu}=0.030$
the entanglement entropy for $\Sigma=0$ is greater than that for $\Sigma=3$. But at ${T\over \mu}=0.020$ which is a relatively low temperature, the entanglement entropy for $\Sigma=0$
becomes less than that for $\Sigma=3$. However, we mention here that for very low temperatures, numerical calculations are not very trustworthy and therefore
we refrain from making any exact statement here.

\begin{figure}[t!]
\begin{minipage}[b]{0.5\linewidth}
\centering
\includegraphics[width=2.7in,height=2.3in]{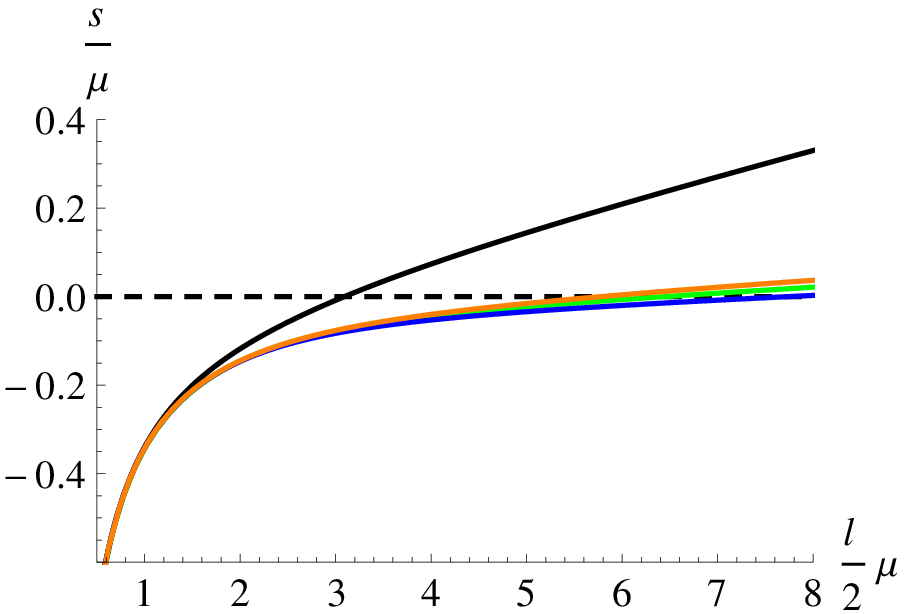}
\caption{HEE for $T=0.5 ~ T_c$, $\Sigma=5$ and $2\kappa^{2}=0.5$ for different values of $\eta$.}
\label{s-lSigma5}
\end{minipage}
\hspace{0.4cm}
\begin{minipage}[b]{0.5\linewidth}
\centering
\includegraphics[width=2.7in,height=2.3in]{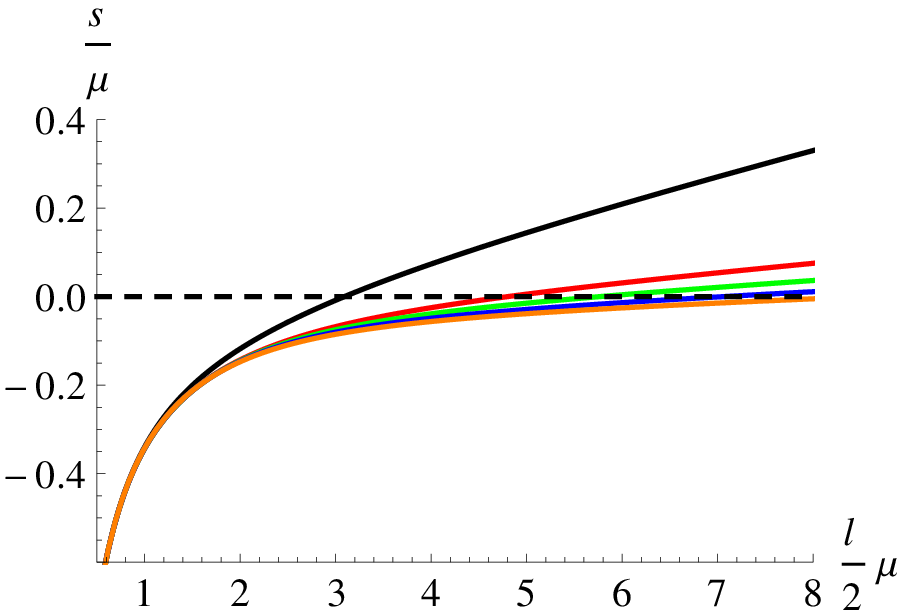}
\caption{HEE for $T=0.5 ~ T_c$, $\eta=-0.1$ and $2\kappa^{2}=0.5$ for different values of $\Sigma$.}
\label{s-leta-0.1}
\end{minipage}
\end{figure}

To complete the analysis, we have also calculated the behavior of the entanglement entropy $s$ as a function of strip width $l$, at a fixed temperature.
This is shown in figs.(\ref{s-lSigma5}) and (\ref{s-leta-0.1}), where we have set $T=0.5 ~ T_c$ and $2\kappa^2=0.5$. The solid black line denotes the HEE for normal phase.
We see that for each case as we increase $l$, $s$ monotonically increases from a negative value and attains a positive value for large $l$.
%%%%%%%%%%%%%%%%%%%%%%%%%%%%%%%%%%%%%%%%%%
%%%%%%%%%%%%%%%%%%%%%%%%%%%%%%%%%%%%%%%%%%

We now turn to the computation of HEE in the VGHS in 4-D R-charged backgrounds, considered in section 2. We take the same metric ansatz with back reaction
as in eq.(\ref{metric4D}) which we reproduce here for convenience
\begin{eqnarray}
\textit{d}s^{2}=-g(r)H(r)^{-1/2}e^{-\chi(r)}dt^{2}+\frac{H(r)^{1/2}}{g(r)}dr^{2}+H(r)^{1/2}r^{2}(dx^{2}+dy^{2})
\label{Rcharged_metric4D}
\end{eqnarray}
Now introducing $z = 1/r$, the above metric can be written as
\begin{eqnarray}
\textit{d}s^{2}=-g(z)H(z)^{-1/2}e^{-\chi(z)}dt^{2}+\frac{H(z)^{1/2}}{z^4
g(z)}dz^{2}+\frac{H(z)^{1/2}}{z^{2}}(dx^{2}+dy^{2})
\label{RchargedZ_metric4D}
\end{eqnarray}
Here, $z=1$ corresponds to the the horizon and $z=0$ to the boundary. We can calculate the HEE for the superconducting and the normal phase in the same
way as we did with the AdS-Schwarzschild background. For this we again take our subsystem $\mathcal{A}$, residing on the
boundary, to be a straight strip and define its domain by $-\frac{l}{2}\leq x \leq \frac{l}{2}$ and $0\leq y \leq L_0$.
Parameterizing the minimal surface $\gamma_\mathcal{A}$ by $x=x(z)$, we first calculate the area of this hypersurface,
\begin{equation}\label{Rcharged_Area}
\mbox{Area}(\gamma_\mathcal{A})=L_0\int_{-l/2}^{l/2}\frac{dx}{z^2}\sqrt{H(z)\Big(1+\frac{z'(x)^2}{z^2
g(z)}\Big)}~.
\end{equation}
When we minimize the above area functional, we get the equation for the minimal surface 
\begin{equation}\label{Rcharged_zstar}
\frac{\sqrt{H(z)}}{z^2\sqrt{1+\frac{z'(x)^2}{z^2 g(z)}}}=\frac{\sqrt{H(z_{*})}}{z_*^2}
\end{equation}
where, as before, $z_{*}$ represents the turning point of the minimal surface such that $z'(x)|_{z=z_{*}}=0$.
Finally, one can write down the entanglement entropy \cite{Ryu} as
\begin{equation}\label{Rcharged_EE}
S={\mbox{Area}(\gamma_\mathcal{A})\over 4 G_4}={2L_0\over
4G_4}\int_{\epsilon}^{z_{*}}dz\frac{z_{*}^2}{z^3}
\frac{H(z)}{\sqrt{(z_{*}^4 H(z)-z^4 H(z_{*}))g(z)}}={2L_0\over
4G_4}(s+\frac{1}{\epsilon})~,
\end{equation}
with
\begin{equation}\label{Rcharged_length}
\frac{l}{2}=\int_{\epsilon}^{z_{*}}dz\frac{z \sqrt{H(z_{*})}}{\sqrt{(z_{*}^4
H(z)-z^4 H(z_{*}))g(z)}} ~
\end{equation}
In eq. (\ref{Rcharged_EE}) the first term $s$ represents the finite part of the EE. Like our previous case with the AdS
Schwarzschild background, we will only concentrate on the computation of the physically relevant finite part $s$ of the EE.

\begin{figure}[t!]
\begin{minipage}[b]{0.5\linewidth}
\centering
\includegraphics[width=2.7in,height=2.3in]{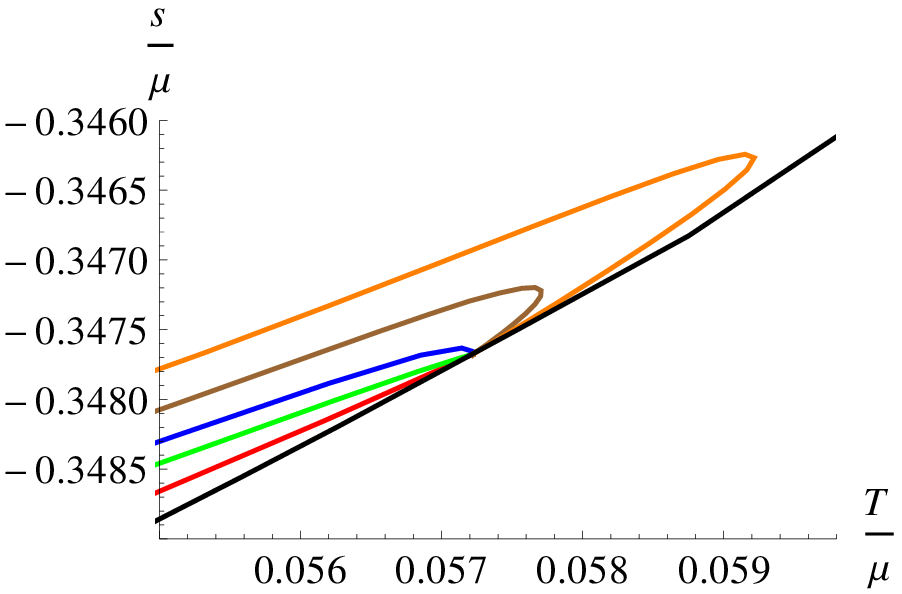}
\caption{HEE for fixed ${l\over 2}\mu=1$, $\eta=-0.1$ and $2\kappa^{2}=0.3$ for
different $\Sigma$.}
\label{Rcharged_s-Teta-0.1}
\end{minipage}
\hspace{0.4cm}
\begin{minipage}[b]{0.5\linewidth}
\centering
\includegraphics[width=2.7in,height=2.3in]{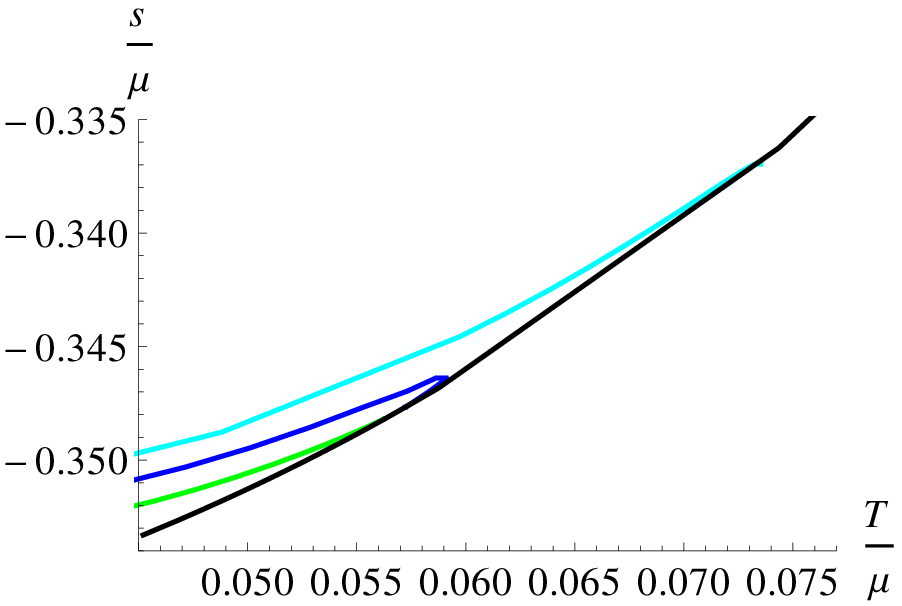}
\caption{HEE for fixed ${l\over 2}\mu=1$, $\Sigma=15$ and $2\kappa^{2}=0.3$ for
different values of $\eta$.}
\label{Rcharged_s-TSigma15}
\end{minipage}
\end{figure}

First we study the behavior of the HEE with temperature, keeping the strip width fixed. We set ${l\over 2}\mu=1$, $\eta=-0.1$,
$2\kappa^2=0.3$ and take different values of $\Sigma$. The results are shown in fig.(\ref{Rcharged_s-Teta-0.1}) where the
black curve denotes the HEE for the normal phase. The curves with red, green, blue, brown and orange color correspond to $\Sigma=1,
5,7,10~\mbox{and}~15$ respectively. The most important observation here is that the HEE in the
superconducting phase is greater than that for the normal phase, which contradicts expected behavior. 
A calculation of the free energy here shows that in the superconducting phase it is smaller than that in the normal phase.
This situation is not repeated in the VGHS in five dimensional R-charged backgrounds, where we find the HEE in the superconducting 
phase is smaller than the normal phase. Therefore, the higher magnitude of the HEE in our set up seems to be a special property of four dimensional R-charged black hole background.

At this point we are unable to explain this  behavior of HEE for the VGHS in four dimensional R-charge backgrounds (as we elaborate shortly). 
However, the order of the phase transition is clear from the figure and it is consistent with our previous result on condensate as
a function of temperature. For $\Sigma=1,5~\mbox{and}~7$, the slope in the HEE shows a discontinuity at the critical
value of ${T\over \mu}$, indicating a second order phase transition.
But as one increases the value of $\Sigma$ from $\Sigma=7$, the HEE becomes multivalued near the critical value of $T/\mu$,
showing a discontinuous jump in $s$, which indicates a first order phase transition. 
%%%%%%%%%%%%%%%%%%%%%%%%%%%%
%%%%%%%%%%%%%%%%%%%%%%%%%%%%
%%%%%%%%%%%%%%%%%%%%%%%%%%%%
\begin{figure}[t!]
\begin{minipage}[b]{0.5\linewidth}
\centering
\includegraphics[width=2.7in,height=2.3in]{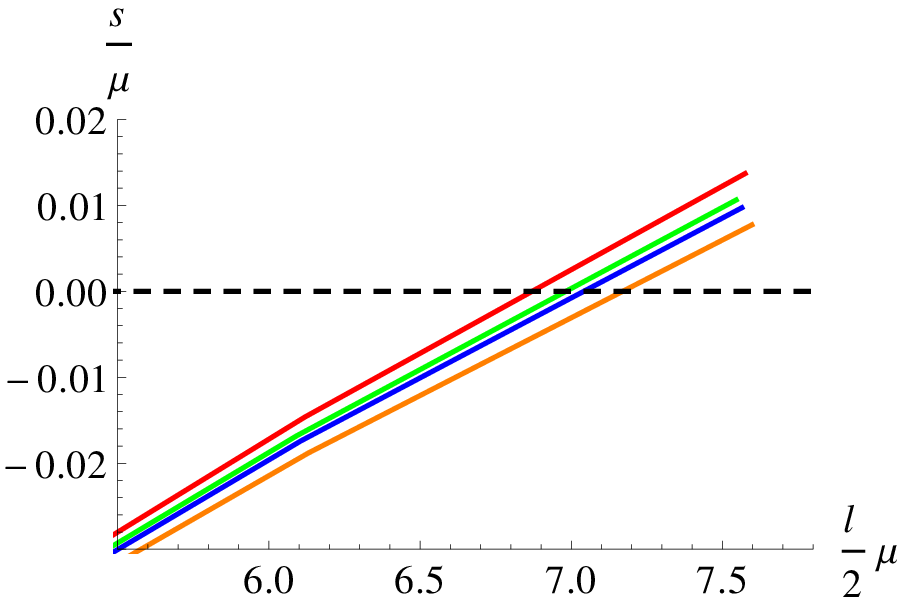}
\caption{HEE for $T=0.5 ~ T_c$, $\eta=-0.1$ and $2\kappa^{2}=0.3$ for different
values of $\Sigma$. Red, green, blue and orange colors correspond to $\Sigma=1,
5,7~\mbox{and}~15$ respectively.}
\label{Rcharged_s-leta-0.1}
\end{minipage}
\hspace{0.4cm}
\begin{minipage}[b]{0.5\linewidth}
\centering
\includegraphics[width=2.7in,height=2.3in]{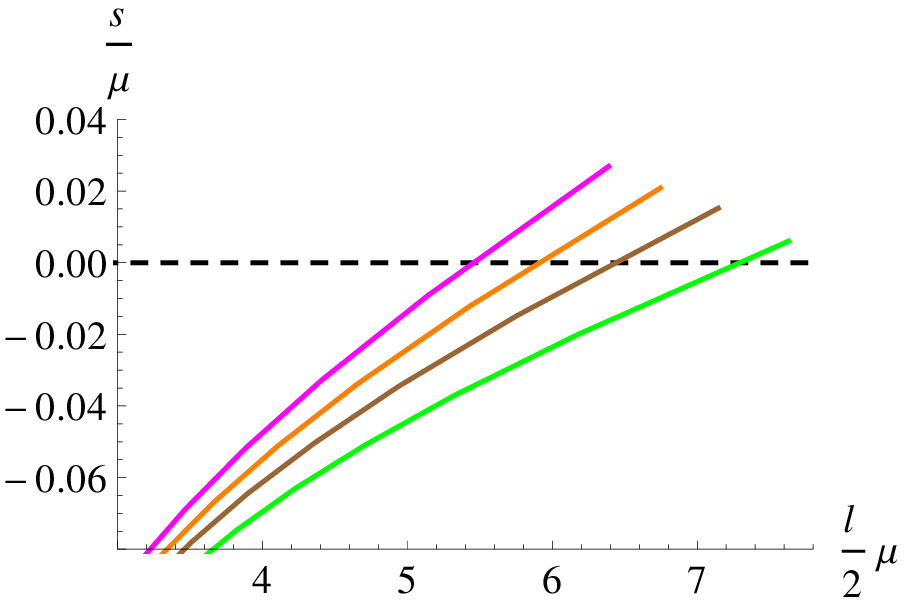}
\caption{HEE for $T=0.5 ~ T_c$, $\Sigma=15$ and $2\kappa^{2}=0.3$ for different
values of $\eta$. Green, brown, orange and magenta correspond to $\eta=-0.01,-0.5,-1~\mbox{and}~-2$
respectively.}
\label{Rcharged_s-lSigma15}
\end{minipage}
\end{figure}
%%%%%%%%%%%%%%%%%%%%%%%%%%%%

Now taking $\eta$ as the varying parameter, and fixing $\Sigma$, we show the HEE in
fig.(\ref{Rcharged_s-TSigma15}), where the black curve represent the HEE in the normal phase. Here we set
${l\over 2}\mu=1$, $\Sigma=15$, $2\kappa^2=0.3$. The curves with green, blue and cyan colors correspond
to $\eta=-0.01,-0.1~\mbox{and}~-3$ respectively. The window of first order transitions should be obvious, 
tut again, the HEE in the superconducting phase seems to be greater than that in the normal phase.

It is difficult to pinpoint the physical reason for this behavior as the computations are entirely numerical. If we set $H=1$ in the 4-D 
R-charged background, we recover the usual behavior for the HEE, as in AdS-Schwarzschild examples. 
Although this would suggest that the difference in the R-charged background is due to the $H(r)$ term in the metric, one has to be careful 
before drawing any conclusion. This is because we have checked that the nature of the functions $g(r)$, $H(r)$ and $\chi(r)$ are all qualitatively similar 
in the VGHS in four as well as five dimensional R-charged backgrounds. The fact that the HEE behaves differently only 
in four dimensions is possibly due to the different nature of the coupled 
differential equations in these systems. We do not have a better understanding of this as of now. 

We have also studied the behavior of entanglement entropy as a function of the strip width $l$ at a fixed temperature.
The results are shown in fig.(\ref{Rcharged_s-leta-0.1}) and (\ref{Rcharged_s-lSigma15}) where we have set $T=0.5 ~ T_c$ and
$2\kappa^2=0.3$. The behavior of the HEE with $l$ is qualitatively similar to that with the AdS-Schwarzschild background we
have just studied. We find that for each case as we increase $l$, $s$ monotonically increases from negative values and attains
a positive values for large $l$.

\section{\textbf{Entangling Temperature of Holographic Superconductors}}

While discussing HEE, it is very interesting to ask whether there exists here a ``first law of thermodynamics.''
Recently, this question has been discussed in \cite{Bhattacharya}, where it is shown that for a small subsystem, the change of the entanglement entropy is
proportional to the change of the energy of the subsystem and the proportionality constant,
which is given by the size of the entangling region, is interpreted as the inverse of the entangling temperature. The procedure to establish this
is to calculate the entanglement entropy and energy for the excited state of a d-dimensional boundary CFT. The dual gravitational picture of this excited state is
the deformed AdS space. Since we want to calculate the entangling temperature in our model of holographic superconductors, we will consider AdS black holes
with scalar hair as the deformed AdS space and as mentioned above this would correspond to excited state of the boundary CFT. Then by calculating the change
in entanglement entropy $(\Delta S)$ and the change in energy $(\Delta E)$ of the boundary CFT due to this deformation, one can calculate the entangling temperature.

The computation of entangling temperature in the context of holography involves a number of steps. We will not mention the details here but refer the interested reader to \cite{Alishahiha}.
As considered in \cite{Bhattacharya}, \cite{Alishahiha}, we choose our ground state in the CFT to be dual to four dimensional pure AdS with metric
\begin{equation}
 \textit{d}s^{2}=\frac{1}{z^2}\left(-dt^2+dz^2+dx^2+dy^2\right)
 \label{pureads}
\end{equation}
and the entanglement entropy of the ground state with a subsystem of straight strip of width $(l)$ is given by

\begin{equation}
S^{(0)}_E=\frac{2 L_0}{4 G_{4}}\bigg{[}\frac{1}{\epsilon}-{2\pi\over l}
\left(\frac{\Gamma\left(\frac{3}{4}\right)}{\Gamma\left(\frac{1}{4}\right)}\right)^{2}
\bigg{]}
\label{GSAdS}
\end{equation}
The excited state of the boundary CFT (the superconducting phase) in our case will be described by the following metric in the bulk,
\begin{eqnarray}
\textit{d}s^{2}={1\over z^2} \left(-f(z)e^{-\chi(z)}dt^{2}+\frac{dz^2}{f(z)}+dx^2+dy^2\right)
\label{defAdS}
\end{eqnarray}
The above metric can be considered as a thermal deformation of the pure AdS geometry (\ref{pureads}) such that the boundary theory which now has a non-zero
temperature corresponds to an excited state. Our strategy here is to compute the form of $f(z)$ and $\chi(z)$ numerically at a
particular temperature (below $T_c$, so that we are in the superconductor phase) and calculate the change in entanglement entropy
caused by this deformation. In order to calculate the change in energy of the subsystem, it is useful to cast the metric (\ref{defAdS}) in Fefferman-Graham
coordinates,
\begin{equation}
\textit{d}s^{2}=\frac{1}{z^2}\bigg(dz^2+g_{\mu\nu} dx^\mu dx^\nu\bigg)
 \label{FG}
\end{equation}
where $g_{\mu\nu}=\eta_{\mu\nu}+h_{\mu\nu}(x,z)$ with
\begin{equation}
h_{\mu\nu}(x,z)=h^{(0)}_{\mu\nu}(x)+z^2 h^{(2)}_{\mu\nu}(x)+z^3 h^{(3)}_{\mu\nu}(x)+\cdots
 \label{FGexp}
\end{equation}
From the expansion it is clear that $h_{\mu\nu}(x,z)$ contains the information about the excited state.

Now at a particular temperature, we fit the numerical solution of $f(z)$ and $\chi(z)$ with the polynomials
\begin{eqnarray}
f(z) &=& 1 +a_3 z^3 +a_4 z^4+\cdots \nonumber\\
\chi(z) &=& A_2z^2 + A_3 z^3 +A_4 z^4+\cdots
\end{eqnarray}
and calculate the coefficients $a_k$ and $A_k$ for all $k$. Note that the form of  $f(z)$ and $\chi(z)$ depend on the higher derivative coupling constant $\eta$, and that the model parameter $\Sigma$,
so the coefficients $a_k$ and $A_k$ will also change accordingly. For example, for the AdS-Schwarzchild black hole background, the polynomial coefficients that fit the curve $f(z)$ and
$\chi(z)$ for $T=0.5T_c$ and $\eta = -0.1$, with the backreaction parameter $2\kappa^2 = 0.5$ are given in tables (1) and (2).\footnote{For ease of presentation,
we have truncated some of the numbers that appear in the following tables. An exact fit obtained by using a standard MATHEMATICA routine provides slightly more precise values.}

%%%%%%%%%%%%%%%%%%%%%%%%%%%%
\begin{center}
\begin{minipage}{\linewidth}
\centering
\captionof{table}{Coefficients of $f(z)$, for fixed $\eta = -0.1$} \label{tab1}
\begin{tabular}{|c|c|c|c|c|c|c|c|}
\hline
$\Sigma$ $\Big\backslash a_i$ &$a_3$ &$a_4$ &$a_5$ &$a_6$ &$a_7$ &$a_8$ &$a_9$\\
\hline
$0$    &-20.261&89.562&-166.732&148.364&-51.171&-7.991&7.229\\
\hline
$1$  &-38.439&220.521&-524.026&639.927&-412.217&123.640&-10.404\\
\hline
$3$  &-76.493&584.037&-1851.370&3124.140&-2967.180&1501.940&-316.077\\
\hline
$5$  &-93.975&781.032&-2660.490&4786.60&-4814.06&2565.35&-565.479\\
\hline
$7$  &-106.590&943.375&-3374.180&6321.33&-6578.09&3608.8&-815.656\\
\hline
\end{tabular}\par
\end{minipage}
\end{center}
\vskip0.5cm
\begin{minipage}{\linewidth}
\centering
\captionof{table}{Coefficients of $\chi(z)$, for fixed $\eta = -0.1$} \label{tab2}
\begin{tabular}{|c|c|c|c|c|c|c|}
\hline
$\Sigma$ $\Big\backslash A_i$ &$A_2$ &$A_3$ &$A_4$ &$A_5$ &$A_6$ &$A_7$\\
\hline
$0$    &-0.830&9.744&-0.151&-31.346&36.772&-12.844\\
\hline
$1$  &-2.790&34.691&-68.335&46.345&-2.647&-5.802\\
\hline
$3$  &-6.319&87.75&-262.584&353.392&-228.937&58.165\\
\hline
$5$  &-7.019&106.638&-345.949&501.759&-348.364&94.374\\
\hline
$7$  &-6.569&117.047&-405.317&617.857&-446.948&125.367\\
\hline
\end{tabular}\par
\bigskip
\end{minipage}
%%%%%%%%%%%%%%%%%%%%%%%%%%%%
\vskip0.2cm
\noindent
Similarly, the coefficients $a_i$ and $A_i$ from polynomial fitting of $f(z)$ and $\chi(z)$ at $T=0.5~T_c$, $2\kappa^2= 0.5$,  and $\Sigma=5$, for different
values of $\eta$, are shown in the following tables (3) and (4) :\\

%%%%%%%%%%%%%%%%%%%%%%%%%%%%
\begin{minipage}{\linewidth}
\centering
\captionof{table}{Coefficients of $f(z)$, for fixed $\Sigma = 5$} \label{tab3}
\begin{tabular}{|c|c|c|c|c|c|c|c|}
\hline
$\eta$ $\Big\backslash a_i$ &$a_3$ &$a_4$ &$a_5$ &$a_6$ &$a_7$ &$a_8$ &$a_9$\\
\hline
$-0.01$    &-48.4672&303.385&-781.153&1051.61&-778.868&296.878&-44.380\\
\hline
$-0.1$  &-93.9752&781.032&-2660.49&4786.6&-4814.06&2565.35&-565.479\\
\hline
$-0.5$  &-72.6527&585.099&-1970.95&3531.54&-3548.66&1892.4&-417.783\\
\hline
$-1$  &-51.9212&373.551&-1157.72&1936.37&-1834.96&930.064&-196.39\\
\hline
\end{tabular}\par
\bigskip
\end{minipage}
\vskip0.5cm
\begin{minipage}{\linewidth}
\centering
\captionof{table}{Coefficients of $\chi(z)$, for fixed $\Sigma = 5$} \label{tab4}
\begin{tabular}{|c|c|c|c|c|c|c|}
\hline
$\eta$ $\Big\backslash A_i$ &$A_2$ &$A_3$ &$A_4$ &$A_5$ &$A_6$ &$A_7$\\
\hline
$-0.01$    &-3.997&49.748&-108.541&92.630&-27.095&-1.053\\
\hline
$-0.1$  &-7.019&106.638&-345.949&501.759&-348.364&94.374\\
\hline
$-0.5$  &-4.188&69.8242&-229.024&334.125&-233.174&63.484\\
\hline
$-1$  &-2.842&43.6855&-131.069&176.526&-114.738&29.323\\
\hline
\end{tabular}\par
\bigskip
\end{minipage}
%%%%%%%%%%%%%%%%%%%%%%%%%%%%

Now by substituting the form of $f(z)$ and $\chi(z)$ into (\ref{defAdS}),
we can cast it into the form of (\ref{FG}) and therefore can calculate the coefficients $h^{(0)}_{\mu\nu}(x)$, $h^{(2)}_{\mu\nu}(x)$,
$h^{(3)}_{\mu\nu}(x)$ etc.
Assuming that $h^{(n)}_{\mu\nu} l^n\ll 1$ throughout our calculation and following \cite{Alishahiha}, we find the increase in entanglement entropy of the excited state with respect to the ground state as
\begin{equation}
 \Delta S_{E}=\frac{1}{4
G_4}\int_0^{z_*}dz\left(\Gamma^{(0)}+\Gamma^{(2)}z^2+\Gamma^{(3)}z^3+\cdots\right)
 \label{delS}
\end{equation}
where
\begin{eqnarray}
\int_\epsilon^{z_*}\Gamma^{(n)}r^n &=& \frac{1}{(1-n)\epsilon^{1-n}} \int dx\;\left( {\rm Tr}(h^{(n)}_{ab})-h^{(n)}_{11}\right) \nonumber \\
&-& F(2,2-n)\frac{l^{n-1}}{2^{n-1}a_\zeta^{n-1}} \int dx\;\left( {\rm Tr}(h^{(n)}_{ab})-\frac{2}{n+1}h^{(n)}_{11}\right)
\end{eqnarray}
with
$$F(m,n)=\frac{_2F_1(\frac{1}{2},\frac{1-n}{2m},\frac{2m+1-n}{2m},1)}{n-1}, \ a_\zeta=\frac{\sqrt{\pi}\Gamma\left(\frac{3}{4}\right)}{\Gamma\left(\frac{1}{4}\right)}$$
Here $_2F_1$ is the Hypergeometric function and $z_*$ is the turning point with pure AdS geometry. Now, using the prescription of \cite{Bala}, \cite{Skenderis}, the energy momentum
tensor of the excited state in the dual CFT side is given as
\begin{equation}
\langle T_{\mu\nu}\rangle=\frac{3}{16\pi G_4}\;h_{\mu\nu}^{(3)}
\end{equation}
We use the above expression to calculate the increase in energy of the excited state as 
\begin{equation}
\Delta E=\int d^2 x \langle \Delta T_{tt}\rangle.
 \label{delE}
\end{equation}
From the above discussion, it is clear that $\Delta E$ is always proportional to $l$. Using (\ref{delS}) and (\ref{delE}), we calculate 
the entangling temperature as $T_{ent}={\Delta E \over \Delta S_E}$.
%%%%%%%%%%%%%%%%%%%%%%%%%%%%%%%%%%%%%%%%%%
\begin{figure}[t!]
\begin{minipage}[b]{0.5\linewidth}
\centering
\includegraphics[width=2.7in,height=2.3in]{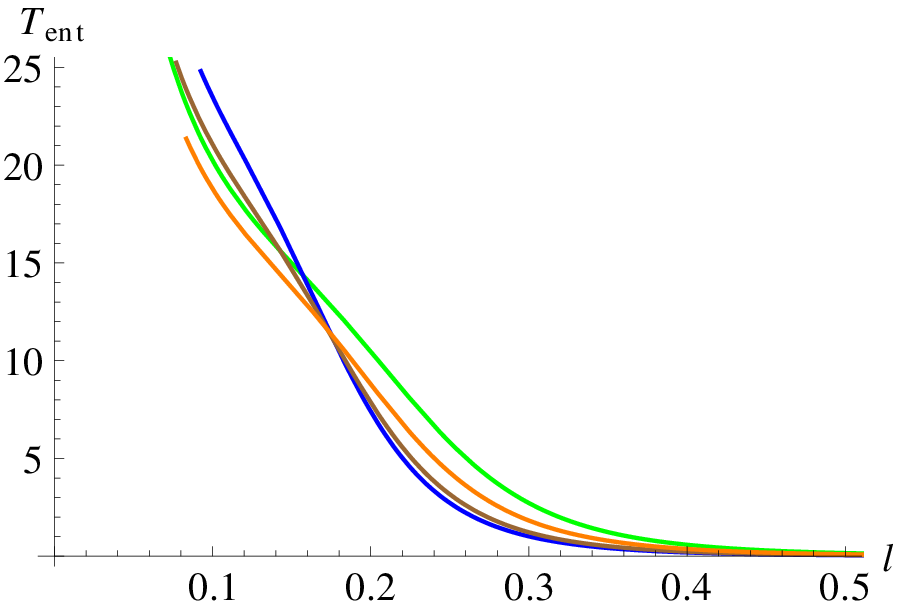}
\caption{$T_{ent}$ vs $l$ at $T=0.5 ~ T_c$, $\Sigma=5$ and $2\kappa^{2}=0.5$ for different values of $\eta$.}
\label{TEnt-Eta}
\end{minipage}
\hspace{0.4cm}
\begin{minipage}[b]{0.5\linewidth}
\centering
\includegraphics[width=2.7in,height=2.3in]{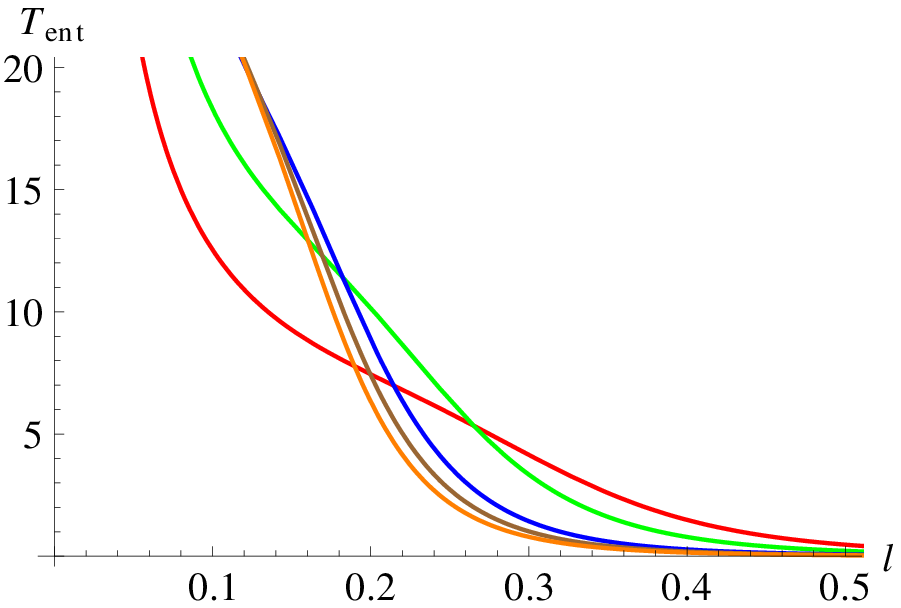}
\caption{$T_{ent}$ vs $l$ at $T=0.5 ~ T_c$, $\eta=-0.1$ and $2\kappa^{2}=0.5$ for different $\Sigma$.}
\label{TEnt-Sigma}
\end{minipage}
\end{figure}
%%%%%%%%%%%%%%%%%%%%%%%%%%%%%%%%%%%%%%%%%%
In fig.(\ref{TEnt-Eta}) we have shown the variation of entangling temperature $T_{ent}$ as a function of the strip width $l$ for different values of $\eta$. The green, blue, brown and orange curves 
correspond to $\eta=-0.01, -0.1, -0.5~\mbox{and}-1$, respectively.
Here we have fixed $T=0.5 ~ T_c$, $\Sigma=5$ and $2\kappa^{2}=0.5$. We note that as one decreases the strip width, $T_{ent}$ increases and diverges as $l\to 0$. Physically,
this corresponds to the fact that at zero strip width, there is no entanglement. 

Qualitatively similar behavior for $T_{ent}$ was observed in \cite{Bhattacharya}, where the authors found $T_{ent}\propto1/l$ for AdS-Schwarzschild black hole (without any scalar hair).
In our case, there are a few differences. First, for the case of the non-hairy AdS-Schwarzschild black hole, only $h^{(3)}_{\mu\nu}(x)$ is non zero and  therefore $T_{ent}$ is always proportional to $1/l$.
For the VGHS, higher order terms in eq.(\ref{FGexp}) can be nonzero and therefore can modify the $T_{ent}\propto1/l$ relation. Indeed this
is what we see in fig.(\ref{TEnt-Eta}).
We can also calculate the departure of $T_{ent}$ from the $1/l$ behavior that appear in the four dimensional non-hairy AdS-Schwarzschild black hole case. This is calculated
as $\frac{T_{ent}^{(1)} - T_{ent}^{(2)}}{T_{ent}^{(1)}}$, where $T_{ent}^{(1)}$ and $T_{ent}^{(2)}$ are the entangling temperatures for the non-hairy AdS-Schwarzschild case and the
VGHS in the AdS-Schwarzschild cases, respectively.
From eqn.(\ref{delS}), one can see that there is a contribution to the entanglement entropy not only from $\Gamma^{(0)}$, $\Gamma^{(2)}$ and
$\Gamma^{(3)}$, but also from the higher order terms. This is because while expanding $h_{\mu\nu}(x)$ in eqn (33), one needs to
consider the terms beyond $h^{(3)}_{\mu\nu}(x)$. The appearance of these extra terms is what modifies the behavior of  $T_{ent}$.
For $\eta=-0.01$, $\Sigma=5$ and $T=0.5 T_c$ we find a departure of around $17\%$ for $l=0.05$ and $37\%$ for $l=0.1$.
However, the departure from $T_{ent}\propto1/l$ is not very large near criticality. For small value of $l$, say
$l\backsim0.25$, we find a departure of around $3\%$ and for $l\backsim0.6$ we find a departure of around $15\%$ near $T_c$ for the same values of $\eta$ and $\Sigma$.

We also extend the analysis for different values of $\Sigma$
for fixed $\eta$ and $\kappa$, at $T=0.5T_c$. This is shown in fig.(\ref{TEnt-Sigma}) where the red, green, blue, brown and orange curves correspond 
to $\Sigma$=$0$, $1$, $3$, $5$ and $7$, respectively.

For the VGHS in R-charge backgrounds, we ran into some difficulties with the above procedure. This is due to the fact that fitting polynomials for $f(z)$ and $\chi(z)$ are
difficult to obtain very precisely. Due to this, the entangling temperature could not be calculated properly. We will not discuss this issue further.

\section{Conclusions and Discussions}

In this concluding section, we will summarize our main results. First, we have constructed a very general class of phenomenological models
for holographic superconductors in single R-charged black hole backgrounds, in four and five dimensions, including back reaction effects. The VGHS models constructed in this paper
correspond to supergravity backgrounds of rotating brane configurations, and hence non-trivially extend the ones considered in \cite{Dey} for the AdS-Schwarzschild case.
We find that our models predict a rich phase structure in the parameter space, with a window of first order phase transitions. \footnote{For AdS-soliton backgrounds appropriate for
studying insulator-superconductor phase transitions, we find that the VGHS does not show any such window.}  As pointed out in the text, this might be
phenomenologically important in the understanding of the strongly coupled behavior of superconductors. In the probe limit, the phase diagram of our model
is qualitatively shown in fig.(\ref{phasediagram}). Admittedly, the results contained in this paper cannot be used to understand realistic physical phenomena as of now,
but these further our understanding of phase transitions in holographic scenarios, and we only hope that they should be useful in future experiments. 

Next, we studied holographic entanglement entropy for our model, and found that the HEE precisely captures the information about the phase transitions alluded to above.
In the window of parameters where a first order phase transition is predicted by a calculation of the free energy, the HEE for the superconducting phase is multi valued,
and is single valued outside. However we find that in the four dimensional example that we have worked out, the HEE seems to be higher in the superconducting
phase, contrary to results that appear in the literature. Since the results are completely numerical, it is difficult to pinpoint the exact reason for this. 

Finally, we studied the entangling temperature for generalized holographic superconductors in the AdS Schwarzschild background. We found that the temperature shows
deviation from a pure AdS background, and that these are dependent on the model parameters. We were unable to perform this calculation in R-charged backgrounds, as
it was difficult to obtain exact fits to the metric components here. This case needs to be further investigated. 

It will be interesting to calculate the optical response properties of the VGHS in R-charge backgrounds, analogous to what was done in \cite{Subhash}, \cite{Dey}. 
It might also be useful to consider different types of higher derivative couplings in holographic models. We leave these issues for a future publication.

\vskip1.2cm
\begin{center}
{\bf Acknowledgements}
\end{center}
The work of SM is supported by grant no. 09/092(0792)-2011-EMR-1 from CSIR, India.
\vskip1cm
\appendix
\section{Details of 5-D single R-charged black hole backgrounds}
In this appendix, we present the details of our calculations for holographic superconductors in 5-D R-charged black hole backgrounds.
We start with the action
\begin{eqnarray}
\textit{S} &=& \int \mathrm{d^{5}}x\! \sqrt{-g}\biggl[\frac{1}{2\kappa^{2}}\biggl(R+4\biggl(H^{2/3}+2H^{-1/3}\biggr)\biggr)
-\frac{H^{4/3}}{8}\textit{F}_{\mu\nu}\textit{F}^{\mu\nu}-\frac{1}{3}\frac{(\partial H)^2}{H^2} \nonumber \\ 
&-&\frac{1}{2}|\textit{D}\tilde{\Psi}|^{2} -\frac{1}{2}m^{2}|\tilde{\Psi}|^{2}
-\frac{\eta}{2}|\textit{F}_{\mu\nu}\textit{D}^{\nu}\tilde{\Psi}|^{2}\biggl] \,
\label{action5D}
\end{eqnarray}
Now writing the charged scalar field as $\tilde{\Psi} =\Psi e^{i\alpha}$ and following section 2, the action of eq.(\ref{action5D}) can be generalized as
\begin{eqnarray}
\textit{S} &=& \int \mathrm{d^{5}}x\! \sqrt{-g}\biggl[\frac{1}{2\kappa^{2}}\biggl(R+4\left(H^{2/3}+2H^{-1/3}\biggr)\right)
-\frac{ H^{4/3}}{8}\textit{F}_{\mu\nu}\textit{F}^{\mu\nu}-\frac{1}{3}\frac{(\partial H)^2}{H^2} \nonumber \\ 
&-&\frac{(\partial_{\mu}\Psi)^2}{2}
-\frac{\eta}{2}\textit{F}_{\mu\nu}\partial^{\nu}\Psi \textit{F}^{\mu\sigma}\partial_{\sigma}\Psi  -\frac{m^2\Psi^2}{2} -\frac{|\textrm{G}(\Psi)|(\partial\alpha-q A)^2}{2} \nonumber \\
&-& \frac{\eta}{2}|\textrm{K}(\Psi)|\biggl(\textit{F}^{\mu\nu}(\partial_{\nu}\alpha-q A_{\nu})\biggr)^{2} \biggr]
\end{eqnarray}
For 5-D background we will consider the following ansatz
\begin{eqnarray}
\textit{d}s^{2}=-g(r)H(r)^{-2/3}e^{-\chi(r)}dt^{2}+H(r)^{1/3}r^{2}(dx^{2}+dy^{2}+dz^{2})+H(r)^{1/3}\frac{dr^{2}}{g(r)}
\label{metric5D}
\end{eqnarray}
Equation of motion for the scalar field $\Psi$
\begin{eqnarray}
&&\Psi '' \left(1-\eta  e^{\chi}H^{1/3} \Phi'^2\right)+ \frac{H e^{\chi} \Phi^2}{2g^2}\frac{d\textrm{G}(\Psi)}{d\Psi} -\frac{\eta H^{4/3} e^{2\chi}\Phi^2 \Phi'^2}{2g^2}
\frac{d\textrm{K}(\Psi)}{d\Psi} \nonumber \\  
&-& \eta H^{1/3} e^{\chi}\Psi'\left(\frac{g' \Phi'^2}{g} + \frac{\chi' \Phi'^2}{2} +\frac{H'\Phi'^{2}}{3H} + \frac{3\Phi'^2}{r} + 2\Phi' \Phi''\right) \nonumber \\ 
&-&\frac{m^2 H^{1/3} \Psi}{g} + \Psi'\left( \frac{3}{r}+\frac{g'}{g}-\frac{\chi '}{2}\right)=0
\label{psieom5D}
\end{eqnarray}
Equation of motion for the zeroth component of the gauge field
\begin{eqnarray}
&& \Phi''\left(1 -\frac{2\eta e^{\chi} \Phi^2 \textrm{K}(\Psi)}{g H^{2/3}} + \frac{2\eta g \Psi'^2}{H^{5/3}} \right) -\Phi \left(\frac{2\textrm{G}(\Psi)}{g H}+\frac{2\eta e^{\chi} \Phi'^2 \textrm{K}(\Psi) }{g H^{2/3}}\right) \nonumber \\ 
&& + \frac{2\eta g \Psi'^2\Phi'}{H^{5/3}} \left(\frac{g'}{g} + \frac{\chi'}{2} + \frac{H'}{3 H} + \frac{3}{r} + \frac{2 \Psi''}{\Psi'}\right) + \Phi'\left(\frac{3}{r} + \frac{2 H'}{H}+\frac{\chi'}{2}\right) \nonumber \\ 
&& + \frac{2\eta e^{\chi} \textrm{K}(\Psi) \Phi ^2 \Phi'}{g H^{2/3}} \left(\frac{g'}{g}-
\frac{\textrm{K}(\Psi)'}{\textrm{K}(\Psi)}-\frac{3\chi'}{2}-\frac{3}{r} - \frac{4 H'}{3H}\right) =0
\label{phieom5D}
\end{eqnarray}
H-filed equation of motion
\begin{equation}
H'' + H' \left(\frac{3}{r}+\frac{g'}{g}-\frac{\chi'}{2}-\frac{H'}{H} \right)+\frac{e^{\chi} H^{3} \Phi'{^2}}{2g}+\frac{4H}{2\kappa^{2}g}(H-1)=0
\label{Heom5D}
\end{equation}
Similarly the Einstein equations give
\begin{eqnarray}
&&g' +\frac{2g}{r}-\frac{4r}{3}(H+2)+\frac{r g' H'}{6 H} + \frac{g H'}{H} - \frac{5r g H'^2}{18 H^2}+ \frac{r g H''}{3 H}  \nonumber \\ 
&&+2\kappa^{2} r \biggl(\frac{H e^{\chi}\Phi^2\textrm{G}(\Psi)}{6 g} + \frac{H^{1/3}m^{2} \Psi^{2}}{6} + \frac{g H'^{2}}{9 H^2} + \frac{\eta g H^{1/3} e^{\chi}\Phi'^{2}\Psi'^{2}}{6} +\frac{e^{\chi}H^{2}\Phi'^{2}}{12} \nonumber \\  
&&- \frac{\eta H^{4/3} e^{2\chi} \Phi^{2}\Phi'^{2}\textrm{K}(\Psi)}{2 g} +\frac{g \Psi'^{2}}{6} \biggr)=0
\label{rreinsteineom5D}
\end{eqnarray}
\begin{eqnarray}
&&2\kappa^{2} r \biggl(\frac{H e^{\chi}\Phi^2\textrm{G}(\Psi)}{3 g^{2}}+\frac{2 H'^{2}}{9H^{2}} -\frac{\eta H^{4/3} e^{2\chi}\Phi^{2}\Phi'^{2}\textrm{K}(\Psi)}{3 g^{2}}+\frac{\Psi'^{2}}{3}
-\frac{\eta H^{1/3} e^{\chi}\Phi'^{2}\Psi'^{2}}{3} \biggr) \nonumber \\ 
&& \chi'\left(1+ \frac{r H'}{6 H}\right)+\frac{H'}{H}-\frac{2rH'{2}}{9H^{2}}+\frac{rH''}{3H}=0
\label{ttrreinsteineom5D}
\end{eqnarray}
here again prime denotes a derivative with respect to r and also r dependence of each variable is suppressed. The hawking temperature for background with the metric (\ref{metric5D}) is given by
\begin{equation}
T_{H}=\frac{g'(r)e^{-\chi(r)/2}}{4\pi \sqrt{H(r)}}|_{r=r_{h}}
\end{equation}
To solve these five coupled differential equations we impose the following boundary conditions
\begin{equation}
\Phi(r_{h})=0 ,\ \ \Psi'(r_{h})=\frac{m^2 H(r_{h})^{1/3}\Psi(r_{h})}{g'(r_{h})(1-\eta H({r_h})^{1/3} e^{\chi(r_{h})}\Phi'^{2}(r_{h}))}.
\label{horizon behavior5D}
\end{equation}
Near the boundary these fields asymptote to the following expressions
\begin{eqnarray}
\Phi=\mu-\frac{\rho}{r^2} +... , ~~ \Psi=\frac{\Psi_{-}}{r^{\lambda_{-}}}+\frac{\Psi_{+}}{r^{\lambda_{+}}} + ... ~~ \chi\rightarrow 0, \ \ \ g\rightarrow r^{2}+..., \ \ H\rightarrow 1+...
\label{boundar behavior5D}
\end{eqnarray}
here $\lambda_{\pm}=\frac{4\pm\sqrt{16+4m^2}}{2}$.

\section{Necessary formulas for the VGHS in 4-D AdS-Schwarzschild backgrounds}

We consider the following action
\begin{eqnarray}
\textit{S} &=& \int \mathrm{d^{4}}x\! \sqrt{-g}\biggl[\frac{1}{2\kappa^{2}}\biggl(R+\frac{6}{L^{2}}\biggr)
-\frac{1}{4}\textit{F}_{\mu\nu}\textit{F}^{\mu\nu}-\frac{1}{2}|\textit{D}_{\mu}\tilde{\Psi}|^{2} \nonumber \\ 
&-&\frac{1}{2}m^{2}|\tilde{\Psi}|^{2} -\frac{\eta}{2}|\textit{F}_{\mu\nu}\textit{D}^{\nu}\tilde{\Psi}|^{2}\biggl] \,
\label{action4D}
\end{eqnarray}
As in section 2, we rewrite
$\tilde{\Psi}= \Psi  e^{i \alpha}$ and the action becomes
\begin{eqnarray}
\textit{S}= \int \mathrm{d^{4}}x\! \sqrt{-g}\biggl[\frac{1}{2\kappa^{2}}\biggl(R+\frac{6}{L^{2}}\biggr)-\frac{1}{4}\textit{F}_{\mu\nu}\textit{F}^{\mu\nu}-\frac{(\partial_{\mu}\Psi)^2}{2}-
\frac{m^2\Psi^2}{2} &\nonumber \\ -\frac{\eta}{2}\textit{F}_{\mu\nu}\partial^{\nu}\Psi \textit{F}^{\mu\sigma}\partial_{\sigma}\Psi -\frac{\Psi^2(\partial\alpha-q A)^2}{2}-
\frac{\eta}{2}\Psi^2\biggl(\textit{F}^{\mu\nu}(\partial_{\nu}\alpha-q A_{\nu})\biggr)^{2} \biggr] \,
\label{actionpsi2a}
\end{eqnarray}
Now we replace $|\Psi|^2$ by two different analytic functions of $\Psi$, ${\rm G}(\Psi)$ and ${\rm K}(\Psi)$, keeping in mind that
the gauge invariance should be preserved. Thus we have our generalized action,
\begin{eqnarray}
\textit{S} &=& \int \mathrm{d^{4}}x\! \sqrt{-g}\biggl[\frac{1}{2\kappa^{2}}\biggl(R+\frac{6}{L^{2}}\biggr)-\frac{1}{4}\textit{F}_{\mu\nu}\textit{F}^{\mu\nu}-\frac{(\partial_{\mu}\Psi)^2}{2}
-\frac{\eta}{2}\textit{F}_{\mu\nu}\partial^{\nu}\Psi \textit{F}^{\mu\sigma}\partial_{\sigma}\Psi \nonumber \\ 
&-& \frac{m^2\Psi^2}{2} -\frac{|\textrm{G}(\Psi)|(\partial\alpha-q A)^2}{2}- \frac{\eta}{2}|\textrm{K}(\Psi)|\biggl(\textit{F}^{\mu\nu}(\partial_{\nu}\alpha-q A_{\nu})\biggr)^{2} \biggr]
\label{action3}
\end{eqnarray}

We take the background metric as,
\begin{eqnarray}
\textit{d}s^{2}=-r^2 f(r)e^{-\chi(r)}dt^{2}+\frac{dr^{2}}{r^2 f(r)}+r^{2}(dx^{2}+dy^{2})
\label{metric4d}
\end{eqnarray}
with the following ansatz
\begin{equation}
\Psi=\Psi(r),~~~A=\Phi(r)dt ~.
\end{equation}
The Hawking temperature of the black hole is given by
\begin{equation}
T_{H}=\frac{r^2 f'(r)e^{-\chi(r)/2}}{4\pi}|_{r=r_{h}}
\label{Hawking}
\end{equation}
where $f(r_{h})=0$ defines the radius of the event horizon, $r_{h}$.

The equations of motion for the scalar field $\Psi(r)$ and the gauge field $\Phi(r)$ are,
\begin{eqnarray}
&\Psi''& \Big(1-\eta \mathrm{e}^{\chi } \Phi'^2\Big) + \Psi '\Big(\frac{4}{r}+\frac{f'}{f}-\frac{\chi '}{2}
 -\frac{\eta \mathrm{e}^{\chi } f' \Phi '^2}{f} -\frac{\eta}{2}  \mathrm{e}^{\chi } \Phi '^2 \chi'
 -\frac{4 \eta  \mathrm{e}^{\chi} \Phi '^2}{r}\nonumber\\
 &-& 2 \eta \mathrm{e}^{\chi }\Phi'\Phi''\Big)
+ \frac{ \Phi ^2 \mathrm{e}^{\chi }}{2r^4 f^2}\frac{d\textrm{G}(\Psi)}{d\Psi}
 -\frac{\eta   \Phi ^2 \mathrm{e}^{2 \chi } \Phi '^2}{2r^4 f^2}\frac{d\textrm{K}(\Psi)}{d\Psi}-\frac{m^2 \Psi }{r^2 f} = 0
\label{scalareom}
\end{eqnarray}
\begin{eqnarray}
&\Phi''& \Big(1+\eta  r^2 f \Psi '^2-\frac{\eta  \textrm{K}(\Psi) \Phi^2 e^{\chi }}{r^2f}\Big)+\Phi'\Big(\eta  r^2 f' \Psi'^2
+\frac{1}{2} \eta r^2 f \chi ' \Psi '^2+4 \eta r f \Psi '^2\nonumber\\
 &+&2 \eta r^2 f \Psi ' \Psi ''+\frac{\chi '}{2}+\frac{2}{r}\Big)
 + \eta \Phi ^2 \Phi'\Big(\frac{\textrm{K}(\Psi) e^{\chi } f'}{r^2 f^2}-\frac{  e^{\chi } \textrm{K}(\Psi)'}{r^2 f}
 -\frac{3 \textrm{K}(\Psi) e^{\chi } \chi '}{2 r^2f}\Big)\nonumber\\
 &-&\Phi \Big(\frac{\textrm{G}(\Psi)}{r^2 f}+\frac{\eta  \textrm{K}(\Psi) e^{\chi } \Phi '(r)^2}{r^2 f}\Big) = 0
 \label{gaugeeom}
\end{eqnarray}
Moreover, the $(t,t)$ and $(r,r)$ components of Einstein equation are
\begin{eqnarray}
&f'&+2\kappa^{2} r \biggl({1\over 4}\frac{\textrm{G}(\Psi) \Phi ^2 e^{\chi }}{r^4 f}
-{3\eta \over 4}\frac{\textrm{K}(\Psi) \Phi ^2 e^{2 \chi } \Phi '^2}{r^4
f}+\frac{1}{4} \eta f e^{\chi } \Phi '^2 \Psi '^2 \nonumber \\
&+& \frac{1}{4} f \Psi '^2 + \frac{1}{4} {m^2 \Psi^2\over r^2}+\frac{1}{4} {e^{\chi } \Phi '^2\over r^2}\biggr)
-{3\over r} +\frac{3f}{r} = 0 
\label{tteinsteineom}
\end{eqnarray}
\begin{equation}
\chi'+2\kappa^{2}r\bigg(\frac{ \textrm{G}(\Psi) \Phi ^2 e^{\chi }}{2 r^4 f^2}
-\frac{ \eta  \textrm{K}(\Psi) \Phi ^2 e^{2 \chi } \Phi '^2}{2 r^4 f^2}
-\frac{1}{2}  \eta  e^{\chi } \Phi'^2 \Psi'^2+\frac{1}{2}\Psi'^2\biggr) = 0
\label{rreinsteineom}
\end{equation}
In the above equations again we have set $q=1$ and chosen the gauge $\alpha=0$. Also, the prime symbol indicates a derivative with respect to $r$.

We solve these four coupled differential equations using appropriate boundary conditions. At $r=r_h$, $\Phi=0$. Near the boundary
they behave as
\begin{eqnarray}
\Phi=\mu-\frac{\rho}{r} +... , ~~ \Psi=\frac{\Psi_{-}}{r^{\lambda_{-}}}+\frac{\Psi_{+}}{r^{\lambda_{+}}} + ..., ~~ \chi\rightarrow 0, \ \ \ g\rightarrow r^{2}+...
\label{boundar behavior}
\end{eqnarray}
We rewrite all the coupled equations in terms of $z=r_h/r$ and
using the above boundary conditions we solve them numerically. Also, we will take the same values of $\xi$, $\theta$ and $\gamma$ as
in \cite{Dey}, i.e, $\xi=0$, $\theta=4$, $\gamma=4$.
At high temperature the condensate will vanish, so $\Psi=0$. The solution becomes
\begin{eqnarray}
\Psi = 0 \,,\qquad\chi=0 \,,\qquad \Phi =  \mu (1-z)\,,
\qquad f = 1 - z^3\left(1+{\kappa^2\mu^2 \over 2}\right)+{z^4\kappa^2\mu^2 \over 2}~
\end{eqnarray}
The temperature is given by 
\begin{eqnarray}
T=\frac{1}{4\pi}(3-{\kappa^2\mu^2 \over 2})~.
\end{eqnarray}

\end{document}